\documentclass[aip,jcp,reprint,groupedaddress,footinbib,citeautoscript]{revtex4-1}

\pdfoutput=1

\usepackage{amsmath}
\usepackage{amssymb} 

\usepackage{graphicx}
\usepackage{dcolumn}
\usepackage{bm}

\usepackage{comment}
\usepackage[utf8]{inputenc}
\usepackage[T1]{fontenc}
\usepackage{mathptmx}
\usepackage{siunitx}
\DeclareSIUnit{\calorie}{cal}
\DeclareSIUnit{\Calorie}{\kilo\calorie}

\usepackage{hyperref}
\hypersetup{colorlinks=true,linkcolor=blue,citecolor=blue,urlcolor=blue}

\usepackage{breakurl}

\usepackage{color}

\newcolumntype{L}[1]{>{\raggedright\arraybackslash}p{#1}}
\newcolumntype{C}[1]{>{\centering\arraybackslash}p{#1}} 
\newcolumntype{R}[1]{>{\raggedleft\arraybackslash}p{#1}} 

\usepackage{braket} 
\newcommand{\ketbra}[2]{\ket{#1}\!\bra{#2}}
\newcommand{\eu}{\mathrm{e}^}
\newcommand{\iu}{\ensuremath{\mathrm{i}}}
\newcommand{\rmd}{\mathrm{d}}
\newcommand{\half}{{\ensuremath{\frac{1}{2}}}}
\newcommand{\thalf}{{\ensuremath{\tfrac{1}{2}}}}

\newcommand{\op}[1]{\ensuremath{\hat{#1}}}
\providecommand{\mat}[1]{\mathsf{#1}}
\renewcommand{\mathbf}[1]{\bm{#1}}
\newcommand{\del}{\nabla}
\DeclareMathOperator{\Tr}{Tr}

\newcommand{\grad}{\del}

\newcommand{\der}[3][]{\frac{\rmd^{#1}{#2}}{\rmd{#3}^{#1}}}
\newcommand{\pder}[3][]{\frac{\partial^{#1}{#2}}{\partial{#3}^{#1}}}
\newcommand{\pders}[3]{\frac{\partial^2{#1}}{\partial{#2}\partial{#3}}}

\newcommand{\eqn}[1]{Eq.\,(\ref{#1})}
\newcommand{\Eqn}[1]{Equation~(\ref{#1})}
\newcommand{\eqs}[1]{Eqs.\,(\ref{#1})}
\newcommand{\fig}[1]{Fig.~\ref{fig:#1}}
\newcommand{\secref}[1]{Sec.~\ref{sec:#1}\@}
\newcommand{\Ref}[1]{Ref.~\onlinecite{#1}}
\newcommand{\Refs}[1]{Refs.~\onlinecite{#1}} 

\begin{document}


\title{Instanton formulation of Fermi's golden rule in the Marcus inverted regime}

\author{Eric R. Heller}
 \email{eric.heller@phys.chem.ethz.ch}
\author{Jeremy O. Richardson}%
 \email{jeremy.richardson@phys.chem.ethz.ch}
\affiliation{ 
Laboratory of Physical Chemistry, ETH Z\"urich, 8093 Z\"urich, Switzerland
}%

\date{\today}

\begin{abstract}
Fermi's golden rule defines
the transition rate
between weakly coupled states
and can thus be used to describe
a multitude of molecular processes including
electron-transfer reactions and light-matter interaction.
However, it can only be calculated if the wave functions of all internal states are known, which is typically not the case in molecular systems.
Marcus theory provides a closed-form expression for the rate constant,
which is a classical limit of the golden rule, 
and indicates the existence of a normal regime and an inverted regime.
Semiclassical instanton theory presents a more accurate approximation to the golden-rule rate
including nuclear quantum effects such as tunnelling,
which has so far been applicable to complex anharmonic systems in the normal regime only.
In this paper we extend the instanton method to the inverted regime and study the properties of the periodic orbit,
which describes the tunnelling mechanism via two imaginary-time trajectories, one of which
now travels in negative imaginary time.
It is known that tunnelling is particularly prevalent in the inverted regime, even at room temperature,
and thus this method is expected to be useful in studying
a wide range of molecular transitions occurring in this regime.
\end{abstract}

\maketitle

\section{Introduction}

Describing chemical reactions which take place on more than just one electronic potential-energy surface
poses one of the primary open challenges in the field of 
chemical reaction dynamics. \cite{Althorpe_2016,Lawrence2019rates}
These processes are relevant to many phenomena which we encounter not only in different disciplines of science, but also in our everyday life.
Ranging from redox reactions to photosynthesis, harvesting light in solar cells, molecular switches and many more, the most fundamental step of these processes is a nonadiabatic transition from one electronic state to another,
leading to a breakdown of the 
Born--Oppenheimer approximation. \cite{Tully2012perspective}

Due to the great interest in these phenomena,
the study of nonadiabatic transitions is 
an important topic for research.
Hence, there exists a plethora of different algorithms to simulate nonadiabatic dynamics,
\cite{Curchod_2018,Tully_2012,Crespo_Otero_2018,Yarkony_2011,Worth_2008,Makri2015QCPI}
from computationally expensive, but accurate methods based on wave-function propagation \cite{Beck_2000,MCTDH,Wang2006flux,Richings2015vMCG} to heuristically motivated, pragmatic methods such as trajectory surface hopping.\cite{Tully_1990,Subotnik_2016,Mai_2018}
Simulating the direct dynamics of a chemical reaction, however, is not usually a practical way to obtain information about the reaction rate, because
the typical time scales of chemical reactions are long.
Instead, a nonadiabatic extension of transition-state theory (TST) is required. 
\cite{Rips1995ET}

The thermal rate constant for the transition from the reactant electronic state, with
internal energy levels $E_0^{\lambda}$ and a partition function $Z_0 = \sum_{\lambda} \mathrm{e}^{-\beta E_0^{\lambda}}$,
to the product electronic state, with 
internal energy levels $E_1^{\nu}$, can be found by applying perturbation theory.
The result to lowest order in the coupling
$\Delta_{\lambda\nu}$ between these states
is given by the famous golden-rule formula \cite{Dirac1927,Wentzel_1927}
generalized for an initial thermal distribution \cite{Zwanzig,Nitzan}
\begin{align}
    \label{equ:k_qs}
    k_{\mathrm{QM}} &= \frac{2\pi}{\hbar} \sum_{\lambda} \frac{\mathrm{e}^{-\beta E_0^{\lambda}}}{Z_0} \sum_{\nu}
    |\Delta_{\lambda\nu}|^2
    \delta(E_0^{\lambda} - E_1^{\nu}) \, ,
\end{align}
whose name (given by Fermi) indicates its overwhelming relevance and applicability in a multitude of different fields,
including nuclear physics, light-matter interactions and nonadiabatic transitions.
\cite{Fermi1974,Schinke_1993,ConicalIntersections2,Parker2005,Ribeiro2018polariton}
One important example of the latter
is the nonadiabatic transfer of an electron from a donor to an acceptor.
\cite{Marcus1993review}

Marcus was awarded with the Nobel prize in chemistry in 1992 \cite{Marcus1993review}
for his work on electron-transfer rate theory. \cite{Marcus1956ET,Marcus1985review}
One of the great triumphs of his theory was the prediction of
an
inverted regime, in which the rates decrease despite increasing thermodynamic driving force, and which was later confirmed by experiment.\cite{Miller1984inverted}
Because of its simplicity and practicability, Marcus theory remains the most commonly applied approach to describe electron-transfer reactions. \cite{Blumberger2015ET,Rosspeintner2013photochemistry,Koslowski2012ET,Pelzer2016,Antoniou2006,Feher1989,LyndenBell2007}

However, there are a number of approximations inherent in Marcus theory,  \cite{ChandlerET}
which includes the assumption of parabolic free-energy curves along the reaction coordinate.
It also employs classical statistics and cannot therefore capture nuclear quantum effects like zero-point energy and quantum tunnelling,
the neglect of which could lead to deviations from the exact rate of several orders of magnitude especially at low temperatures.
The inclusion of these effects in novel nonadiabatic rate theories
which can be applied to molecular systems without making the parabolic approximation
is therefore a major objective. \cite{Althorpe_2016}

In particular it has been predicted that quantum tunnelling effects can substantially speed up the rate in the inverted regime.\cite{Siders1981inverted}
This was confirmed experimentally in \Ref{Akesson1991,*Akesson1992},
in which 
reaction rates were found to be up to 8 orders of magnitude larger than were predicted by 
Marcus theory,
but which could be explained by including a quantum-mechanical treatment of the vibrational modes. \cite{Jortner1988}
Early approaches such as these for including quantum statistics into Marcus theory
were, however, only possible by
restricting the system to simplistic models such as the spin-boson model. \cite{Siders1981quantum}
On the other hand, classical golden-rule transition-state theory \cite{nonoscillatory,ChandlerET} has no restriction on the complexity of the system,
but does not take account of quantum nuclear effects.

A domain of methods which proved to be particularly successful in describing nuclear quantum effects in more general systems is based on Feynman's path-integral formulation of quantum mechanics.\cite{Feynman}
For instance, Fermi's golden rule has been recast into a semiclassical instanton formulation,\cite{GoldenGreens}
which does not require detailed knowledge about the internal states and can therefore be applied to complex systems.
However, there is a problem with these methods in the inverted regime because the imaginary-time propagator, on which most of these methods rely, diverges.
Hence, many previous semiclassical
\cite{Wolynes1987dissipation,Cao1995nonadiabatic,*Cao1997nonadiabatic,*Schwieters1998diabatic,Ranya2019instanton,GoldenGreens,GoldenRPI,AsymSysBath}
and imaginary-time path-integral approaches \cite{Wolynes1987nonadiabatic,Schwieters1999diabatic,Lawrence2019ET}
did not tackle the inverted regime.

One approach for extending these methods to treat
the inverted regime was suggested by Lawrence and Manolopoulos, \cite{Lawrence2018Wolynes} who analytically continued Wolynes' rate expression \cite{Wolynes1987nonadiabatic} into the inverted regime.
The rate is then obtained by extrapolating the path-integral data collected in the normal regime into the inverted regime.
While their methodology appears to work very well at predicting rates,
the mechanistic view may be lost by this approach,
as the rate is not extracted directly from a simulation of the system in the inverted regime.

The electron-transfer rate in the inverted regime
cannot be tackled directly by standard ring-polymer molecular dynamics \cite{Habershon2013RPMDreview}
where the transferred electron is treated as an explicit particle 
\cite{Menzeleev2011ET}
as can be explained by a semiclassical analysis. \cite{Shushkov2013instanton}
However, more recent modifications of ring-polymer molecular dynamics, \cite{Menzeleev2014kinetic,*Kretchmer2016KCRPMD,Duke2016Faraday,Tao2019RPSH}
and golden-rule quantum transition-state theory (GR-QTST) \cite{GRQTST,GRQTST2}
have started to address this problem,
but these still lack the rigour, simplicity of implementation and mechanistic insight of instanton theory.

In this paper, we propose an extension of the semiclassical instanton method
\cite{GoldenGreens,GoldenRPI,AsymSysBath}
for the inverted regime
in the golden-rule limit. 
The rate expression in the inverted regime is derived by analytic continuation of the formula in the normal regime, leading to a one-shot method which requires no extrapolation
of results collected in a different regime.
We show excellent agreement with the exact golden-rule rates for the model systems studied.
At the same time it gives direct mechanistic insights, as it automatically locates the dominant tunnelling pathway for the reaction under investigation, which is equivalent to predicting the mechanism.
It can therefore be used to shed light on the role of quantum nuclear tunnelling in electron-transfer reactions,
which is expected to be of substantial importance, especially in the inverted regime.

The instanton approach can be implemented using a ring-polymer discretization,
in which the only change necessary to make the algorithm applicable for the inverted regime is a slight variation in the optimization scheme,
which
turns out to be just as reliable and effective as the optimization in the normal regime.
Hence, the method is conceptually ideally suited for use in conjunction with \textit{ab-initio} potentials and thereby for realistic simulations of molecular systems,
as has been demonstrated for the standard ring-polymer instanton approach. \cite{porphycene,i-wat2,Asgeirsson2018instanton,hexamerprism,Rommel2012enzyme}


\section{Instanton theory in the normal regime}

In this section we summarize our previous derivation of semiclassical instanton theory
for the golden-rule rate in the normal regime.\cite{GoldenGreens}
This follows a similar approach to our derivation of instanton theory on a single Born--Oppenheimer potential. \cite{AdiabaticGreens,InstReview}

We consider a general multidimensional system with two electronic states,
each with a nuclear Hamiltonian of the form
\begin{align}
	\label{Hn}
	\op{H}_n &= \sum_{j=1}^D \frac{\op{p}_j^2}{2m} + V_n(\op{\mat{x}}), 
\end{align}
where
$n\in\{0,1\}$ is the electronic-state index
and
$\mat{x}=(x_1,\dots,x_D)$
are the Cartesian coordinates 
of $D$ nuclear degrees of freedom.
These nuclei move on the potential-energy surface $V_n(\mat{x})$
with conjugate momenta $\op{p}_j$.
Without loss of generality, the nuclear degrees of freedom have been mass-weighted such that each has the same mass, $m$.
The electronic states $\ket{n}$ are coupled by $\Delta(\hat{\mat{x}})$ to give the total Hamiltonian
in the diabatic representation, \cite{ChandlerET}
\begin{align}
	\op{H} &= \op{H}_0 \ketbra{0}{0} + \op{H}_1 \ketbra{1}{1} + \Delta(\op{\mat{x}}) \big( \ketbra{0}{1} + \ketbra{1}{0} \big).
\end{align}
We shall take the diabatic coupling to be constant, $\Delta(\op{\mat{x}}) = \Delta$, and 
assume that it is very weak, i.e.\ $\Delta\rightarrow0$, known as the golden-rule limit,
which is typically the case in electron-transfer reactions. \cite{ChandlerET}
The quantum-mechanical rate is then given by the golden-rule expression, \eqn{equ:k_qs},
which is valid both in the normal and inverted regimes.
However, in order to calculate the rate in this way, 
the internal states of both the reactant and product would be required,
which are typically not known and cannot be computed for a complex system.

\subsection{Correlation function formalism}

The purpose of the semiclassical instanton approach is to obtain a good approximation to the golden-rule rate
without detailed knowledge of the internal states.
Therefore, instead of using the expression in \eqn{equ:k_qs},
we employ the alternative exact definition of the quantum rate \cite{Miller1983rate,ChandlerET,nonoscillatory}
\begin{equation}
    k_{\mathrm{QM}} Z_0 = \frac{\Delta^2}{\hbar^2} \int_{-\infty}^{\infty} c(\tau+\iu t) \,\rmd t \, ,
    \label{equ:ExactK}
\end{equation}
where the flux correlation function is
\begin{equation}
    c(\tau+\iu t) = \Tr \left[ \mathrm{e}^{-\hat{H}_0 (\beta\hbar - \tau-\mathrm{i} t  )/\hbar}\mathrm{e}^{-\hat{H}_1 (\tau + \mathrm{i} t )/\hbar}\right] \, ,
    \label{equ:fcf}
\end{equation}
and the reactant partition function is $Z_0 = \Tr\big[ \mathrm{e}^{-\beta \hat{H}_0} \big]$.
Note that in order to write the expression in this form, it is necessary to assume that the energies of the internal states of both the reactant and product are bounded from below, i.e.\ there exists a finite-energy ground state of $\op{H}_0$ and $\op{H}_1$. 
We shall in addition assume that the energies are not bounded from above, which is the typical situation for molecular Hamiltonians.

In this section we shall choose 
$\tau$ in the range $0<\tau<\beta\hbar$.
Under these circumstances,
it can be shown that $c(\iu z)$ is an analytic function of $z=t-\iu\tau$ 
and as such the integral $\int_{-\infty}^\infty c(\iu z) \, \rmd z$ is independent of the contour of integration
and hence the rate is independent of the choice of $\tau$, at least within its range of validity.

Expanding the trace in a coordinate-space representation gives
\begin{multline}
    k_{\mathrm{QM}} Z_0 = \frac{\Delta^2}{\hbar} \iiint_{-\infty}^{\infty} K_0(\mat{x}',\mat{x}'',\beta\hbar - \tau - \mathrm{i}t )\\
    \times K_1(\mat{x}'',\mat{x}',\tau + \mathrm{i}t)  \, \rmd \mat{x}' \, \rmd \mat{x}'' \,\rmd t ,
    \label{equ:pi_rate}
\end{multline}
where
the imaginary-time quantum propagators, defined by
\begin{equation}
    \label{equ:kernel}
    K_n(\mat{x}_\text{i},\mat{x}_\text{f},\tau_n) = \braket{\mat{x}_\text{f} | \mathrm{e}^{-\tau_n\hat{H}_n/\hbar} |\mat{x}_\text{i}} \, ,
\end{equation}
describe the dynamics of the system evolving from the initial position $\mat{x}_\text{i}$ to the final position $\mat{x}_\text{f}$
in imaginary time $\tau_n$
according to the Hamiltonian $\hat{H}_n$.
Real-time dynamics can also be described by 
making the third argument complex.

\Eqn{equ:pi_rate} is valid for systems in both the normal and inverted regimes and can be evaluated for model systems where the propagators are known analytically by numerical integration.
However, because it is necessary to limit ourselves to the range $0 < \tau < \beta\hbar$,
as we will show,
the semiclassical approximation
described in \secref{SC} 
can only be derived directly for the normal regime.

\subsection{Semiclassical approximation}
\label{sec:SC}

The instanton expression for the rate is obtained by first replacing the exact quantum propagators by semiclassical van-Vleck propagators  \cite{GutzwillerBook} generalized for imaginary-time arguments \cite{Miller1971density,InstReview}
\begin{equation}
    K_n(\mat{x}_\text{i},\mat{x}_\text{f},\tau_n) \sim \sqrt{\frac{C_n}{(2\pi\hbar)^D}} \mathrm{e}^{- S_n/\hbar} \, .
\end{equation}
This expression is evaluated using the classical trajectory, $\mat{x}_n(u)$,
which travels
from $\mat{x}_n(0) = \mat{x}_\text{i}$ to $\mat{x}_n(\tau_n) = \mat{x}_\text{f}$
in imaginary time $\tau_n$.
This trajectory is found as the path
which makes the Euclidean action, $S_n$, stationary,
and the action is defined as
\begin{align}
    \nonumber S_n &= S_n(\mat{x}_\text{i}, \mat{x}_\text{f}, \tau_n)\\
    &= \int_{0}^{\tau_n} \left[ \frac{1}{2} m \| \dot{\mat{x}}_n(u) \|^2 + V_n(\mat{x_n}(u)) \right] \rmd u \, ,
    \label{equ:action}
\end{align}
where $\dot{\mat{x}}_n(u) = \frac{\rmd \mat{x}_n}{\rmd u}$ is the imaginary-time velocity.
The prefactor of the semiclassical propagator is given by the determinant
\begin{equation}
C_n = \left| -\frac{\partial^2S_n}{\partial \mat{x}_\text{i} \partial \mat{x}_\text{f}} \right| \, .
\label{equ:C_n}
\end{equation}

Plugging this semiclassical propagator into \eqn{equ:pi_rate}
allows us to perform
the integrals over the
end-points $\mat{x}'$, $\mat{x}''$ and over time $t$
employing the method of steepest descent. \cite{BenderBook}
This leads to the following expression for the golden-rule instanton rate, \cite{GoldenGreens}
\begin{equation}
    k_{\mathrm{SCI}} Z_0 = \sqrt{2\pi\hbar} \frac{\Delta^2}{\hbar^2} \sqrt{\frac{C_0 C_1}{- \Sigma}} \mathrm{e}^{-S/\hbar} ,
    \label{equ:kinst}
\end{equation}
where the total action is
\begin{equation}
    S = S(\mat{x}',\mat{x}'',\tau) = S_0(\mat{x}',\mat{x}'',\beta\hbar-\tau) + S_1(\mat{x}'',\mat{x}',\tau) ,
    \label{equ:total_action}
\end{equation}
and the determinant arising from the steepest-descent integration is
\begin{equation}
\Sigma = \begin{vmatrix}
\frac{\partial^2S}{\partial \mat{x}' \partial \mat{x}'} & \frac{\partial^2S}{\partial \mat{x}' \partial \mat{x}''}  & \frac{\partial^2S}{\partial \mat{x}' \partial \tau}  \\ 
\frac{\partial^2S}{\partial \mat{x}'' \partial \mat{x}'}  & \frac{\partial^2S}{\partial \mat{x}'' \partial \mat{x}''}  & \frac{\partial^2S}{\partial \mat{x}'' \partial \tau} \\ 
\frac{\partial^2S}{ \partial \tau \partial \mat{x}'} & \frac{\partial^2S}{ \partial \tau \partial \mat{x}''} & \frac{\partial^2S}{\partial \tau^2}
\end{vmatrix} \, .
\end{equation}
All these expressions are evaluated at a set of values for $\mat{x}'$, $\mat{x}''$ and $\tau$
which describes the stationary point of the action 
defined by $\frac{\partial S}{\partial\mat{x}'} = \frac{\partial S}{\partial\mat{x}''} = \frac{\partial S}{\partial \tau} = 0$.
If $\tau$ is chosen according to this prescription, it is not even necessary to deform the integration contour,
as the saddle point appears at $t=0$ with both $\mat{x}'$ and $\mat{x}''$ purely real.
The minus sign before $\Sigma$ in \eqn{equ:kinst} arises naturally from the Cauchy--Riemann relations \cite{ComplexVariables}
when re-expressing derivatives with respect to $t$ as derivatives with respect to $\tau$.
If there is more than one stationary point of the action,
the rate is given by a sum over each solution. \cite{GRQTST2}
For consistency, the reactant partition function, $Z_0$, should also be evaluated within a semiclassical approximation.\cite{InstReview}

If we instead take the steepest-descent integration over the end-points first and then separately over time,
this leads to the alternative, but equivalent expression
\footnote{This is effectively the same result as was obtained in \Ref{Cao1997nonadiabatic} from an analysis of the ``barrier partition function,'' 
although 
in that paper, the extra approximation that $C_0C_1\approx CZ_0^2$ was also made,
which is exact for the spin-boson model but not in general}
\begin{equation}
    k_{\mathrm{SCI}} Z_0 = \sqrt{2\pi\hbar} \frac{\Delta^2}{\hbar^2} \sqrt{\frac{C_0 C_1}{C}} \left( - \frac{\rmd^2 S}{\rmd\tau^2} \right)^{-\frac{1}{2}} \mathrm{e}^{-S/\hbar}
   \label{equ:kinst2} \, ,
\end{equation}
where
\begin{equation}
C = \begin{vmatrix}
\frac{\partial^2S}{\partial \mat{x}' \partial \mat{x}'} & \frac{\partial^2S}{\partial \mat{x}' \partial \mat{x}''}\\
\frac{\partial^2S}{\partial \mat{x}'' \partial \mat{x}'}  & \frac{\partial^2S}{\partial \mat{x}'' \partial \mat{x}''}
\end{vmatrix}.
\end{equation}

Thus the total action, $S(\mat{x}',\mat{x}'',\tau)$ [\eqn{equ:total_action}], is a sum of the actions of two imaginary-time classical trajectories, one for each electronic state.
One trajectory travels on the reactant state from $\mat{x}'$ to  $\mat{x}''$ in imaginary time $\tau_0 = \beta\hbar-\tau$
and the other from $\mat{x}''$ to $\mat{x}'$ in imaginary time $\tau_1 = \tau$ on the product state.
This forms a closed path of total imaginary time $\tau_0 + \tau_1 \equiv \beta\hbar$, known as the instanton.

Classical trajectories in imaginary time are described by ordinary classical mechanics but with an upside-down potential. \cite{Miller1971density}
They describe quantum tunnelling by travelling through the classically forbidden region,
and typically ``bounce'' against the potential,
which we define as an encounter with a turning point such that the momentum is instantaneously zero. \cite{GoldenGreens}

As has been shown in \Ref{GoldenGreens} for instantons in the normal regime,
the fact that $\mat{x}'$ and $\mat{x}''$ are chosen as stationary points of $S$
is tantamount to saying that
the imaginary-time
momenta on each surface, $\mat{p}_n(u)=m\dot{\mat{x}}_n(u)$, 
must have the same magnitude and point in the same direction at the end-points.
Hence, the two classical trajectories join smoothly into each other.
Furthermore the restriction $\pder{S}{\tau}=0$ ensures 
energy conservation,
which implies that the instanton is a periodic orbit. 
Typically the two end-points will be located in the same place,
which we call the
hopping point, $\mat{x}^\ddag$,
and
we can conclude that 
it
must be located on the crossing seam of the two potentials, where $V_0(\mat{x}^\ddag) = V_1(\mat{x}^\ddag)$.
Further details about these statements will be given in \secref{analysis}.

All the steps in this derivation are asymptotic approximations, which become exact in the $\hbar\rightarrow0$ limit (with $\beta\hbar$ kept constant).
This is therefore known as a semiclassical approximation.
Semiclassical instanton theory gives the exact rate for a system of two linear potentials,
and for more general systems in the limit of high temperature or heavy masses
it approaches a harmonic approximation to classical transition-state theory. \cite{GoldenGreens}
In practice, the theory is applied using a ring-polymer discretization of the imaginary-time trajectories
following the approach described in \Ref{GoldenRPI}.
This method has been previously used to study tunnelling effects and
the dependence of asymmetrical reorganization energies in a system-bath model in the normal regime. \cite{AsymSysBath}

\subsection{Classical limit}
\label{sec:classical}

Here we consider the classical, high-temperature limit ($\beta\hbar\rightarrow 0$) of a general curve crossing problem.
The instanton for this system corresponds to two very short imaginary-time trajectories describing a periodic orbit which is collapsed onto an infinitesimally short line. \cite{GoldenGreens}
Note that unlike for instanton theory on a single Born--Oppenheimer potential, \cite{Miller1975semiclassical,Perspective}
there is no cross-over temperature for golden-rule instanton theory.

For a one-dimensional system in this classical limit,
the action of a single trajectory can be written in the simpler form
\begin{equation}
    \label{Sclassical}
    S_n(x_\text{i},x_\text{f},\tau_n) = \frac{m}{2\tau_n} x_-^2 + V_n(x_+) \tau_n \,,
\end{equation}
where $x_- = x_\text{f} - x_\text{i}$ and $x_+ = \tfrac{1}{2} (x_\text{i} + x_\text{f})$.
The stationary points of the total action, \eqn{equ:total_action},
can be found by searching first for the solution to 
$\pder{S}{x_-}=0$, which gives $x_-=0$,
and then for the solution to $\pder{S}{\tau}=0$ evaluated at $x_-=0$, which requires that the hopping point, $x_+=x^\ddag$,
obeys $V_0(x^\ddag)=V_1(x^\ddag)$.
Finally $\pder{S}{x_+}=0$ requires that
\begin{align}
    \label{equ:tau}
    \tau=\beta\hbar \frac{\grad V_0(x^\ddag)}{\grad V_0(x^\ddag) - \grad V_1(x^\ddag)},
\end{align}
where $\grad V_n(x^{\ddagger})$ is the derivative of the potential $V_n(x)$ with respect to $x$ evaluated at $x^{\ddagger}$.
These solutions give the simple interpretation that the transition-state for the reaction is located at the crossing point between the two diabatic potentials, $x^\ddag$.
Although the value of $\tau$ which makes $S$ stationary does not have a clear interpretation within the classical theory,
it plays an important role in defining the instanton.
We shall therefore consider the behaviour of $\tau$ in various regimes.

The common definition of the ``inverted'' regime is typically expressed in the context of Marcus theory
by a system which has a larger driving force than reorganization energy. 
A more general definition is
that the different regimes are defined by the slope of the potentials at the crossing point;
in the inverted regime the gradients 
have the same sign,
whereas in the normal regime the gradients have opposite signs. 
An alternative terminology for these two cases 
is ``Landau--Zener'' type or ``nonadiabatic-tunnelling'' type. \cite{NakamuraNonadiabatic}
Note that the common definition is equivalent to the more general definition as long as
the driving force is defined as $\varepsilon=V_0(x_\text{min}^{(0)})-V_1(x_\text{min}^{(1)})$
and the (product) reorganization energy as $\Lambda=V_1(x_\text{min}^{(0)})-V_1(x_\text{min}^{(1)})$,
where $x_\text{min}^{(n)}$ is the minimum of $V_n(x)$.
In multidimensional systems, there is a crossing seam, and one would say that the scalar product of the two gradient vectors on this seam is positive only in the inverted regime.
In fact at the minimum-energy crossing point, 
which is the location of the hopping point in the classical limit,
these gradient vectors are antiparallel in the normal regime \cite{GoldenGreens} but parallel in the inverted regime.

From \eqn{equ:tau} it can be seen that in the normal regime,
where $\grad V_0(x^\ddag)$ and $\grad V_1(x^\ddag)$ have opposite signs,
the value of $\tau$ which makes $S$ stationary falls in the range
$0 < \tau < \beta\hbar$ and is therefore always positive.
In the activationless regime, where $\grad V_0(x^\ddag) = 0$, the situation changes to $\tau = 0$.
In this paper, we shall
consider only one type of inverted regime,
$\grad V_1(x^\ddag) / \grad V_0(x^\ddag) > 1$,
which is the typical case encountered in Marcus theory.
Here $\tau$ takes a negative value.
Needless to say, all our results could be easily converted
to describe a system in the alternative inverted regime, $\grad V_0(x^\ddag) / \grad V_1(x^\ddag) > 1$.
\Eqn{equ:tau} generalizes to multidimensional systems by 
projecting each gradient vector along the same arbitrary direction,
and $\tau$ is thus seen to have the same behaviour.

Therefore, in the context of the high-temperature limit of a general (anharmonic) system of two potential-energy surfaces which cross, 
we have shown that %
the sign of $\tau$ is different in the two regimes.
This rule is also known to hold for situations involving quantum tunnelling, 
\cite{Weiss,Lawrence2018Wolynes,nonoscillatory}%
$^,$
\footnote{The argument can easily be extended for the archetypal one-dimensional crossed linear system defined by
$V_n(x) = \kappa_n (x-x^\ddag)$, for which
$S_n = m x_-^2/2\tau_n - \kappa_n (x_+-x^\ddag) \tau_n - \kappa_n^2 \tau_n^3/ 24 m$.
Solving for the stationary point 
$\pder{S}{x_+}=0$ gives $\tau=\beta\hbar \kappa_0 / (\kappa_0 - \kappa_1)$, which is in agreement with \eqn{equ:tau}.
The other two equations give $x_-=0$ and $x_+=x^\ddag$ as expected
}
which will be confirmed by the analysis later in this paper.

When substituting 
the action [\eqn{Sclassical}] evaluated at its stationary point
into \eqn{equ:kinst}, one obtains the classical rate
\begin{equation}
    \label{clTST}
    k_{\text{cl-TST}} Z_0^\text{cl} = \sqrt{\frac{2\pi m }{\beta\hbar^2}}
    \frac{\Delta^2}{\hbar |\grad V_0(x^\ddagger) - \grad V_1(x^\ddagger)|}
    \, \mathrm{e}^{-\beta V^\ddagger} \, ,
\end{equation}
where $V^\ddag = V_0(x^\ddag) = V_1(x^\ddag)$ and $Z_0^\text{cl}$ is the classical limit of the reactant partition function.
Note that this expression for the rate is equivalent to that
derived from classical statistical mechanics employing the Landau--Zener hopping probability in the golden-rule limit.\cite{Nitzan,nonoscillatory}
It can also be derived directly from classical golden-rule TST, \cite{nonoscillatory,ChandlerET}
which is proportional to $\int \eu{-\beta V_0}\delta(V_0-V_1)\,\rmd x$, by noting that
the integral can be performed easily for one-dimensional systems due to the constraint.
For a spin-boson model, this classical expression reduces to Marcus theory.

\Eqn{clTST}
is in fact valid, not just in the normal regime, but also for the inverted regime.
That is
whether $\grad V_0(x^\ddag)$ and $\grad V_1(x^\ddag)$ have the same sign or opposite signs, as long as they are not equal to each other.
It is noteworthy that we can obtain valid classical formulas for the inverted regime from this approach
even though in the derivation we assumed that $0<\tau<\beta\hbar$, which is only appropriate in the normal regime.
In \secref{inverted} we shall also attempt to generalize the instanton method to the inverted regime in a similar way.


\section{Instanton theory in the inverted regime}
\label{sec:inverted}

In \secref{classical} we defined the inverted regime using the gradients on the crossing seam
and found that the value of $\tau$ which makes $S$ stationary becomes negative.
With the definitions given above this
implies $\tau_1<0$ and $\tau_0 > \beta\hbar$, in such a way that $\tau_0+\tau_1 \equiv \beta\hbar$ still holds.
This has important consequences for the implementation and interpretation of the instanton approach,
which we discuss in this section.

\subsection{Analytic continuation}

The main complication for the derivation of instanton theory in the inverted regime
is that the expression for the
imaginary-time propagator $K_1$
diverges for negative $\tau_1$.
One cannot therefore write the rate in terms of the coordinate-space integral as in \eqn{equ:pi_rate}
using the appropriate value for $\tau$, which would be necessary in order to carry out the steepest-descent integration.
However, we will show that while this path-integral expression does diverge,
Fermi's golden rule remains well defined and can be approximated to good accuracy by an
analytic continuation of the instanton rate formula.

Here we study more carefully the cause of the divergence in the inverted regime.
To do this,
we shall investigate the correlation function written as a
sum over 
the contributions of the eigenstates,
$\psi_0^{\lambda}$, of the reactant Hamiltonian, $\op{H}_0$, labelled by the quantum number  $\lambda$
and the eigenstates,
$\psi_1^{\nu}$, of the product Hamiltonian, $\op{H}_1$, labelled by $\nu$.
\footnote{In order for the rate to exist, we assume that at least one of the reactant or product has quasi-continuous states.  If the states are truly continuous, one should replace the sums by integrals.
}
The flux correlation function, \eqn{equ:fcf}, expanded in the energy basis is thus given by
\begin{subequations}
    \label{equ:trace2}
\begin{align}
    \tilde{c}(\tau) &= \sum_{\nu}  g_{\nu}(\tau) \, ,\\
    g_\nu(\tau) &= \sum_{\lambda} |\theta_{\lambda\nu}|^2 \mathrm{e}^{-(\beta\hbar-\tau) E^{\lambda}_0/\hbar - \tau E^{\nu}_1 /\hbar}\, ,
\end{align}
\end{subequations}
where $\theta_{\lambda\nu} = \int \psi_0^{\lambda}(\mat{x})^* \psi_1^{\nu}(\mat{x}) \, \rmd \mat{x}$ such that
the coupling between states used in \eqn{equ:k_qs} is given by $\Delta_{\lambda\nu}=\Delta \theta_{\lambda\nu}$.
The overlap integral is clearly bounded by $0 \le |\theta_{\lambda\nu}| \le 1$, which follows from the Cauchy--Schwarz inequality assuming the wave functions are square-integrable and normalized.

In order to discuss the convergence of the correlation function,
let us first assume that we have a finite system for which the partition functions exist at any temperature,
i.e.\ the sums
$\sum_\nu \eu{-\tau_n E_n^\nu/\hbar}$ converge for any $\tau_n>0$.
In both the normal and inverted regime we are interested in values of $\tau<\beta\hbar$,
and so
it is clear from the comparison test 
that the sum over $\lambda$ converges making $g_{\nu}(\tau)$ a well-defined quantity in either regime.
By similar arguments,
the sum over $\nu$ is also clearly convergent
when $\tau$ is in the range $0<\tau<\beta\hbar$,
which would be appropriate only for the normal regime.

Let us also consider the 
flux correlation function expanded in the position basis,
\begin{equation}
    c(\tau) = \iint K_0(\mat{x}',\mat{x}'',\beta\hbar - \tau ) K_1(\mat{x}'',\mat{x}',\tau) \, \rmd \mat{x}' \, \rmd \mat{x}'' \, ,
\end{equation}
where $K_n(\mat{x}_\text{i},\mat{x}_\text{f},\tau_n) = \sum_{\nu} \psi_n^{\nu}(\mat{x}_\text{i})^* \psi_n^\nu(\mat{x}_\text{f}) \eu{-\tau_n E_n^\nu/\hbar}$.
We expect that for a physical system, $|\psi_n^\nu(\mat{x})|$ 
is bounded by a positive number, $\Psi_\text{max}$, for all $\nu$ and $\mat{x}$.
Therefore, the absolute value of any term in the sum 
is necessarily less than
$\Psi_\text{max}^2 \eu{-\tau_n E_n^\nu/\hbar}$.
The comparison test
(also known as the ``Weierstrass M test'') \cite{ComplexVariables}
can again be invoked to prove that the propagators converge for $\tau_n>0$
for any values of $\mat{x}'$ and $\mat{x}''$,
which is known as uniform convergence. \cite{ComplexVariables}
Note however that $K_1(\mat{x}_\text{i},\mat{x}_\text{f},\tau_1)$ diverges in the inverted regime
if we choose $\tau_1<0$, as is required to perform the steepest-descent integration.

Inserting the definition for $\theta_{\lambda\nu}$ into $\tilde{c}(\tau)$ and the wave-function expansion for $K_n$ into $c(\tau)$, it is clear that 
the two correlation functions are defined in a similar way, the only difference being
that the sums and integrals are taken in a different order.
Only for a uniformly convergent series can we interchange sums and integrals without affecting the result,
and thus it is possible to show that $\tilde{c}(\tau) = c(\tau)$ only for $0<\tau<\beta\hbar$. 
Because $\tilde{c}(\iu z)$ and $c(\iu z)$ are analytic functions of $z=t-\iu\tau$ in the regions where they converge,
this analysis can easily be extended to study the correlation functions at any value of $t$.
This simply adds phases to each term which
does not change the convergence behaviour and the fact that they are identical 
for $0<\tau<\beta\hbar$.

\begin{figure}
    \includegraphics[width=8.5cm]{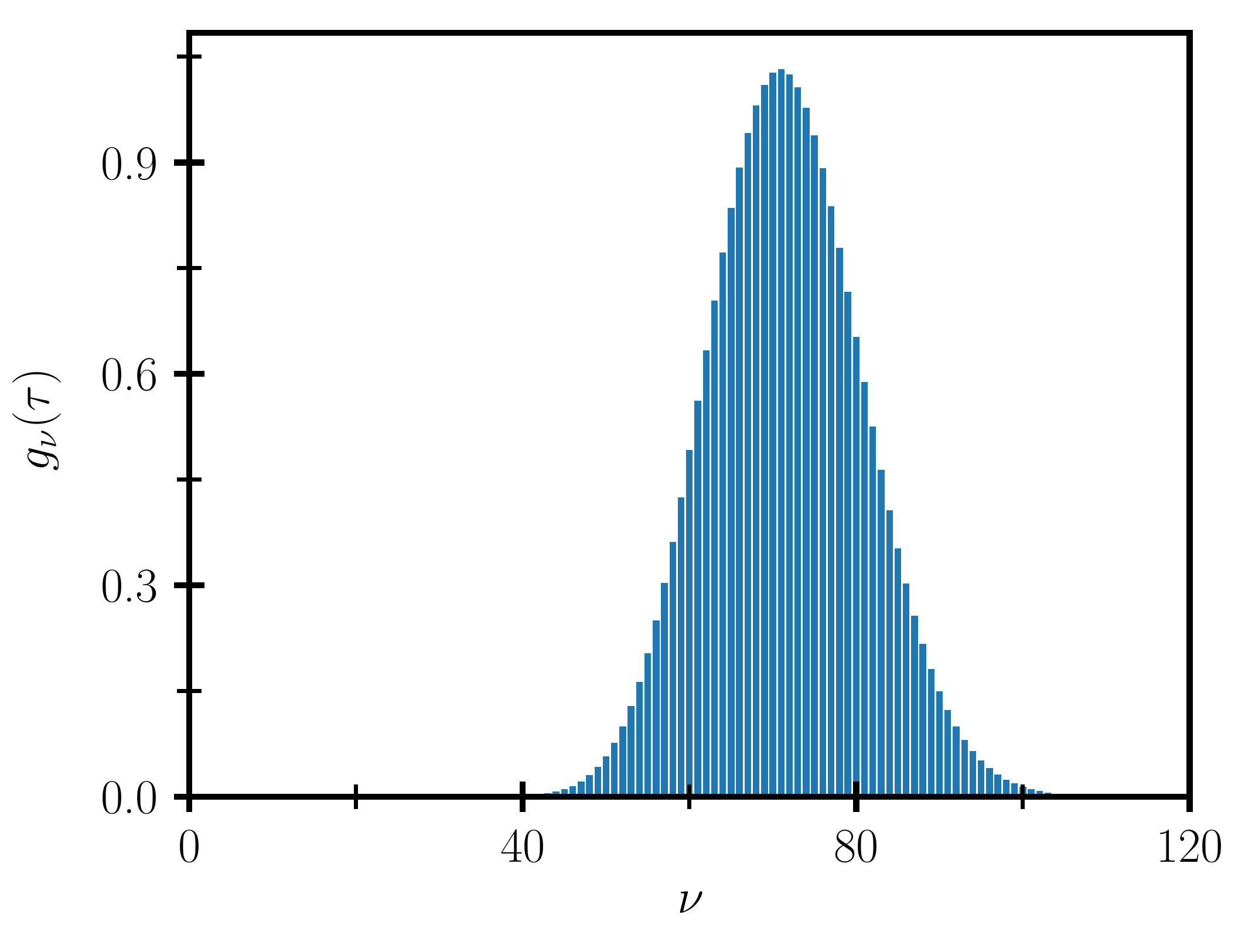}
    \caption{Terms contributing to the trace defined by \eqn{equ:trace2}
    each of which corresponds to a product eigenstate with quantum number $\nu$.
    The terms are computed for the spin-boson model of Sec.~\ref{subsec:SB} with $D=1$ and $\varepsilon/\Lambda=2$
    for $\tau\approx-0.26\beta\hbar$, which is the stationary point of the action for this system.
    }
    \label{fig:traces}
\end{figure}

If we choose $\tau$ to be negative, however, as would be appropriate for the case of the inverted regime,
the sum in $K_1$ is no longer uniformly convergent
and thus $c(\tau)$ diverges.
Interestingly, we find that
the correlation function $\tilde{c}(\tau)$ 
remains
(at least in some cases)
well defined. 
To demonstrate this, we take as an example the one-dimensional version of the spin-boson model defined in Sec.~\ref{subsec:SB}
deep in the inverted regime 
with 
a driving force twice that of the reorganization energy.
Using the value of $\tau$ which makes $S$ stationary,
we evaluate the terms in \eqn{equ:trace2}
in the eigenfunction bases of the two harmonic oscillators.
In Fig.~\ref{fig:traces},
we plot the contributions from each term in the series
and demonstrate
that the $g_{\nu}(\tau)$ terms exhibit a distinct peak at some value of $\nu$ and fall off rapidly either side.
This occurs because $\theta_{\lambda\nu}$ exponentially decreases for states of widely different energies.
Therefore the correlation function $\tilde{c}(\tau)$ converges in this example and is analytic
even for systems in
the inverted regime where $\tau$ is negative. 

In the normal regime,
we know how to make good semiclassical approximations to $c(\tau)$ using instanton theory but have no simple approach based on $\tilde{c}(\tau)$.
Therefore, in order to formulate instanton theory in the inverted regime,
we employ the mathematical process of analytic continuation. \cite{ComplexVariables}
This allows us to evaluate an approximation to $c(\iu z)$ for positive $\tau$,
which because $c(\iu z)=\tilde{c}(\iu z)$,
must also be a good approximation to $\tilde{c}(\iu z)$ in this regime.
Because $\tilde{c}(\iu z)$ is analytic across both regimes, this approximation will be valid also in the inverted regime.
Accordingly, we propose to analytically continue the instanton method into the inverted regime
and will
employ the semiclassical instanton rate expression of \eqn{equ:kinst}, 
not just in the normal regime, where it was originally derived, but also for the inverted regime.

Note the important distinction of this proposed approach to previous work.
In effect, the method of \Ref{Lawrence2018Wolynes} analytically continued the function $c(\tau)$ into the region with $\tau<0$ by fitting it numerically to a suitable functional form \cite{Lawrence2019rates}
based on calculations in the normal regime and extrapolating to describe systems in the inverted regime.
We will go one step further to find a semiclassical instanton approach which is directly applicable in the inverted regime and requires no extrapolation.
In the following we analyse this new approach and show that it gives a valid approximation to Fermi's golden-rule rate.

\subsection{Analysis of the inverted-regime instanton orbit}
\label{sec:analysis}

Through analytic continuation, we have a formula [\eqn{equ:kinst}] for the golden-rule rate in the inverted regime
based on an action $S(\mat{x}',\mat{x}'',\tau)$.
This should be evaluated at its stationary point, which defines the instanton and in this regime has a negative value for $\tau$.
In this section, we shall study the behaviour of the instanton in the inverted regime
and establish that it remains a periodic orbit which travels through classically forbidden regions.

We start with the imaginary-time Euclidean action of a single trajectory, $S_n(\mat{x}_\text{i},\mat{x}_\text{f},\tau_n)$, defined by \eqn{equ:action},
and will be careful to ensure that all our formulas are valid for both positive and negative imaginary times, $\tau_n$.
The trajectory of interest, $\mat{x}_n(u)$, 
starts at 
$\mat{x}_n(0)=\mat{x}_\text{i}$ and travels to $\mat{x}_n(\tau_n)=\mat{x}_\text{f}$ in imaginary time $\tau_n$.
This trajectory has a conserved energy, $E_n$, because the Hamiltonian is time-independent.
We can therefore add a zero under the integral
\begin{subequations}
\begin{multline}
    S_n(\mat{x}_\text{i},\mat{x}_\text{f},\tau_n) = \int_{0}^{\tau_n} \Big[ \thalf m \|\dot{\mat{x}}_n(u)\|^2 + V_n(\mat{x}_n(u))\\ 
    - E(u) + E_n \Big] \rmd u \, ,
    \label{equ:action_E}
\end{multline}
where $E(u) = -\frac{1}{2} m \| \dot{\mat{x}}_n(u) \|^2 + V_n(\mat{x}_n(u))$ is the instantaneous energy,
which is constant (independent of $u$) and equal to $E_n$.
Inserting this definition in \eqn{equ:action_E} leads to
\begin{align}
    \label{equ:legendre_1}
    S_n(\mat{x}_\text{i},\mat{x}_\text{f},\tau_n) &= \int_0^{\tau_n} \left[ m \|\dot{\mat{x}}_n(u)\|^2 + E_n \right] \rmd u \\
     \label{equ:legendre}
      &= \int_{\mat{x}_\text{i}}^{\mat{x}_\text{f}} \mat{p}_n \cdot \rmd \mat{x}_n + E_n\tau_n \, ,
\end{align}
\end{subequations}
where
$\rmd\mat{x}_n$ is an infinitesimal displacement vector pointing along the path in the direction from $\mat{x}_\text{i}$ to $\mat{x}_\text{f}$.
We call this the direction of particle flow.
Our convention will be to define the imaginary-time momentum, $\mat{p}_n=m\der{\mat{x}}{u}$,
such that it points along the
direction of change of position in a positive imaginary-time interval.
Therefore for a trajectory travelling from $\mat{x}_\text{i}$ to $\mat{x}_\text{f}$ in positive imaginary time $\tau_n>0$,
the momentum, $\mat{p}_n(u)$, will point along the path in the direction from $\mat{x}_\text{i}$ to $\mat{x}_\text{f}$.
However, for a trajectory travelling $\mat{x}_\text{i}$ to $\mat{x}_\text{f}$ in negative imaginary time $\tau_n<0$,
the momentum, $\mat{p}_n(u)$, will point in the opposite direction, i.e.\ along the path in the direction from $\mat{x}_\text{f}$ to $\mat{x}_\text{i}$.

From these equations we can determine that
$S_n(\mat{x}_\text{i},\mat{x}_\text{f},\tau_n) \equiv - S_n(\mat{x}_\text{i},\mat{x}_\text{f},-\tau_n)$.
This can be seen from \eqn{equ:legendre_1} or \eqn{equ:action}
as the integral changes sign when the integration range goes from 0 to a negative number.
Alternatively one can see that both terms in \eqn{equ:legendre} change sign when $\tau_n<0$
because 
for negative imaginary times
the momentum vector, $\mat{p}_n$, points in the opposite direction from the particle flow,
i.e.\ it is antiparallel to $\rmd \mat{x}_n$.
In particular if the zero potential is chosen below the instanton energy
(for instance at the reactant minimum),
making $E_n\ge0$,
then $S_n$ will be positive when $\tau_n>0$ and negative when $\tau_n<0$.
Therefore, whereas $S_0$ remains positive just as in the normal regime, in the inverted regime the value of $S_1$ becomes negative. 

As the instanton corresponds to a stationary point of the total action,
we need to know the derivatives of the individual actions of each trajectory
with respect to the initial and final end-points as well as with respect to imaginary time.
These can be found by taking derivatives of \eqn{equ:legendre} to give
\begin{subequations}
\begin{align}
    \pder{S_n(\mat{x}_\text{i},\mat{x}_\text{f},\tau_n)}{\mat{x}_\text{i}} &= -\mat{p}_n(0)\, , \\
    \pder{S_n(\mat{x}_\text{i},\mat{x}_\text{f},\tau_n)}{\mat{x}_\text{f}} &= \mat{p}_n(\tau_n)\, , \\\
    \pder{S_n(\mat{x}_\text{i},\mat{x}_\text{f},\tau_n)}{\tau_n} &= E_n \, .
\end{align}
\end{subequations}
Note that the derivative with respect to the  
initial point has a different sign from that with respect to the 
final point. \cite{GutzwillerBook}

By utilizing these relations in the definition of the total action of the closed orbit, \eqn{equ:total_action}, we arrive at the conditions
\begin{subequations}
\label{equ:conditions}
\begin{align}
    \pder{S}{\mat{x}'} &= -\mat{p}'_0 + \mat{p}'_1 \, ,\\
    \pder{S}{\mat{x}''} &= \mat{p}''_0 - \mat{p}''_1 \, ,\\
    \pder{S}{\tau} &= -E_0 + E_1 \, ,
\end{align}
\end{subequations}
where
$\mat{p}_n'$ is the momentum of the trajectory $\mat{x}_n$ at the end marked $\mat{x}'$
(and likewise for double primes),
i.e.\ $\mat{p}_0' = m\dot{\mat{x}}_0(0)$, $\mat{p}_0'' = m\dot{\mat{x}}_0(\tau_0)$, $\mat{p}_1' = m\dot{\mat{x}}_1(\tau_1)$, $\mat{p}_1'' = m\dot{\mat{x}}_1(0)$.
All of the derivatives in \eqs{equ:conditions} must simultaneously vanish at the instanton configuration,
which effectively imposes energy and momentum conservation at the intersection of the trajectories, 
$\mat{p}_0'=\mat{p}_1'$, $\mat{p}_0''=\mat{p}_1''$ and $E_0=E_1$.
The simplest solution to these equations (and typically the only one) is found when
both $\mat{x}'$ and $\mat{x}''$ are located at the same coordinate, which we call the hopping point, $\mat{x}^\ddag$. 
The first two conditions
require that at the hopping point 
the momenta $\mat{p}_0$ and $\mat{p}_1$ must be vectors with the same direction and the same magnitude. 
Because the third condition requires the energies 
of trajectory to match, $E_0=E_1$, and to be conserved along the path, the potentials at the hopping point must be identical as well.
We can thus conclude that 
the hopping point is located somewhere on the crossing seam between the two potential-energy surfaces
such that $V_0(\mat{x}^\ddag) = V_1(\mat{x}^\ddag)$.
These findings are 
equivalent to those found in previous work limited to the normal regime \cite{GoldenGreens}
but we have now shown that they also hold in the inverted regime.
However, there is nonetheless a fundamental difference
in the inverted regime when we study the paths followed by the trajectories
due to the fact that one path travels in negative imaginary time.

\begin{figure} 
\includegraphics[width=8.5cm]{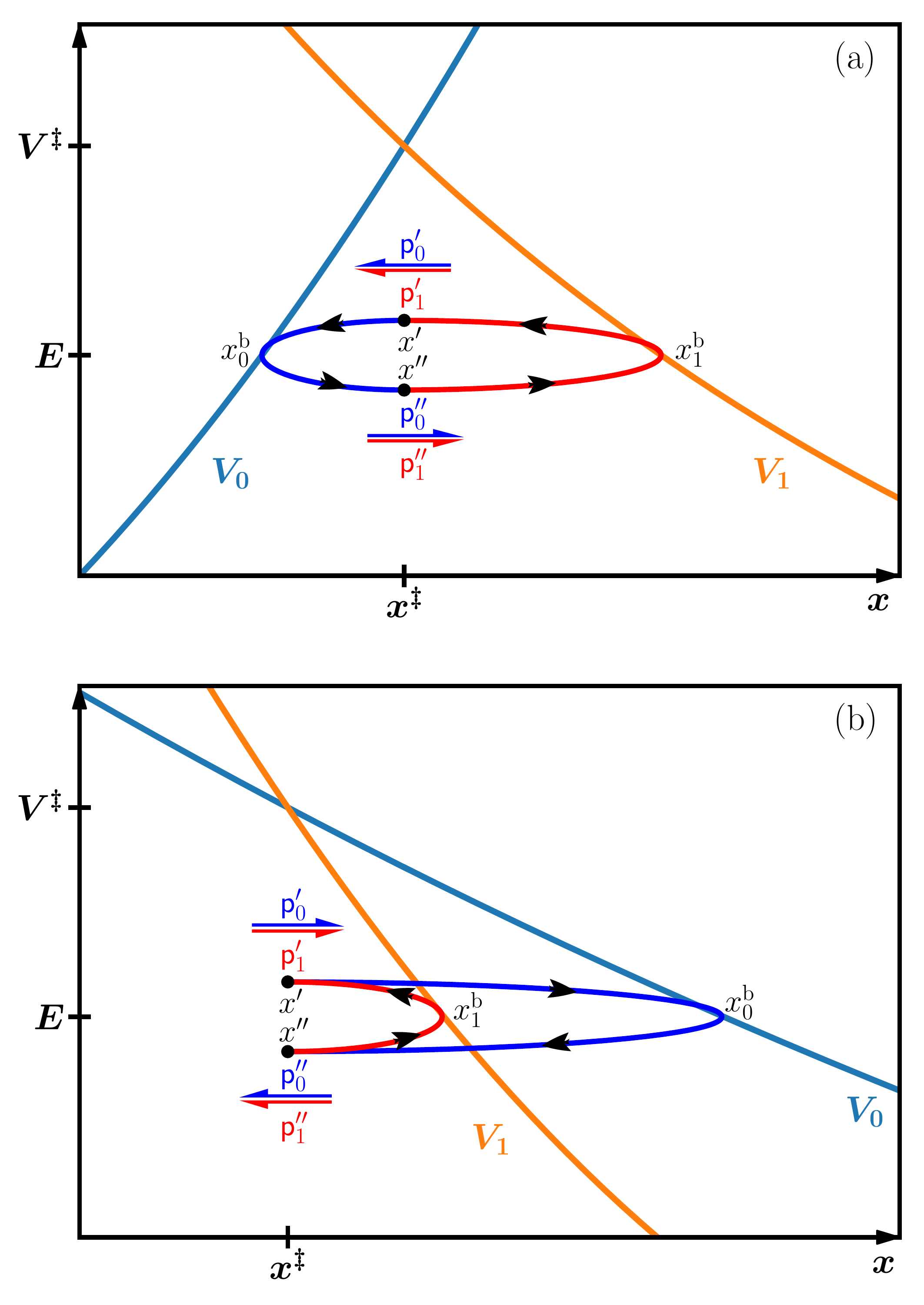}
\caption{Visualization of the two imaginary-time trajectories forming instantons in (a) the normal and (b) the inverted regime at energy $E=E_0=E_1$.
The reactant trajectory, $x_0$, is given in blue and the product trajectory, $x_1$, in red.
Arrows indicate the direction 
of particle flow from the initial point to the final point of each trajectory.
The steepest-descent integration of positions is taken about the crossing point $x'=x''=x^{\ddagger}$ at which $V_0(x^{\ddagger}) = V_1(x^{\ddagger}) = V^{\ddagger}$.
In the normal regime, 
the trajectories bounce on either side of the crossing point,
i.e.\ $x_0^\text{b} < x^\ddag < x_1^\text{b}$,
whereas in the inverted regime,
both trajectories bounce on the same side of the crossing seam, i.e.\ $x^\ddag < x_1^\text{b} < x_0^\text{b}$.
}
\label{fig:instvis}
\end{figure}

The imaginary-time classical trajectories, $\mat{x}_n(u)$,
which start and end at the same point $\mat{x}^\ddag$ but travel in a non-zero amount of time $\tau_n$, whether positive or negative,
will typically bounce against the potential.
This happens halfway along at the turning point, $\mat{x}_n^\text{b} = \mat{x}_n(\tau_n/2)$,
at which the kinetic energy is zero and the total energy, $E_n=V_n(\mat{x}_n^\text{b})$.
These considerations, along with the conditions in \eqn{equ:conditions}, give us the picture shown in Fig.~\ref{fig:instvis}.

\begin{figure} 
\includegraphics[width=8.5cm]{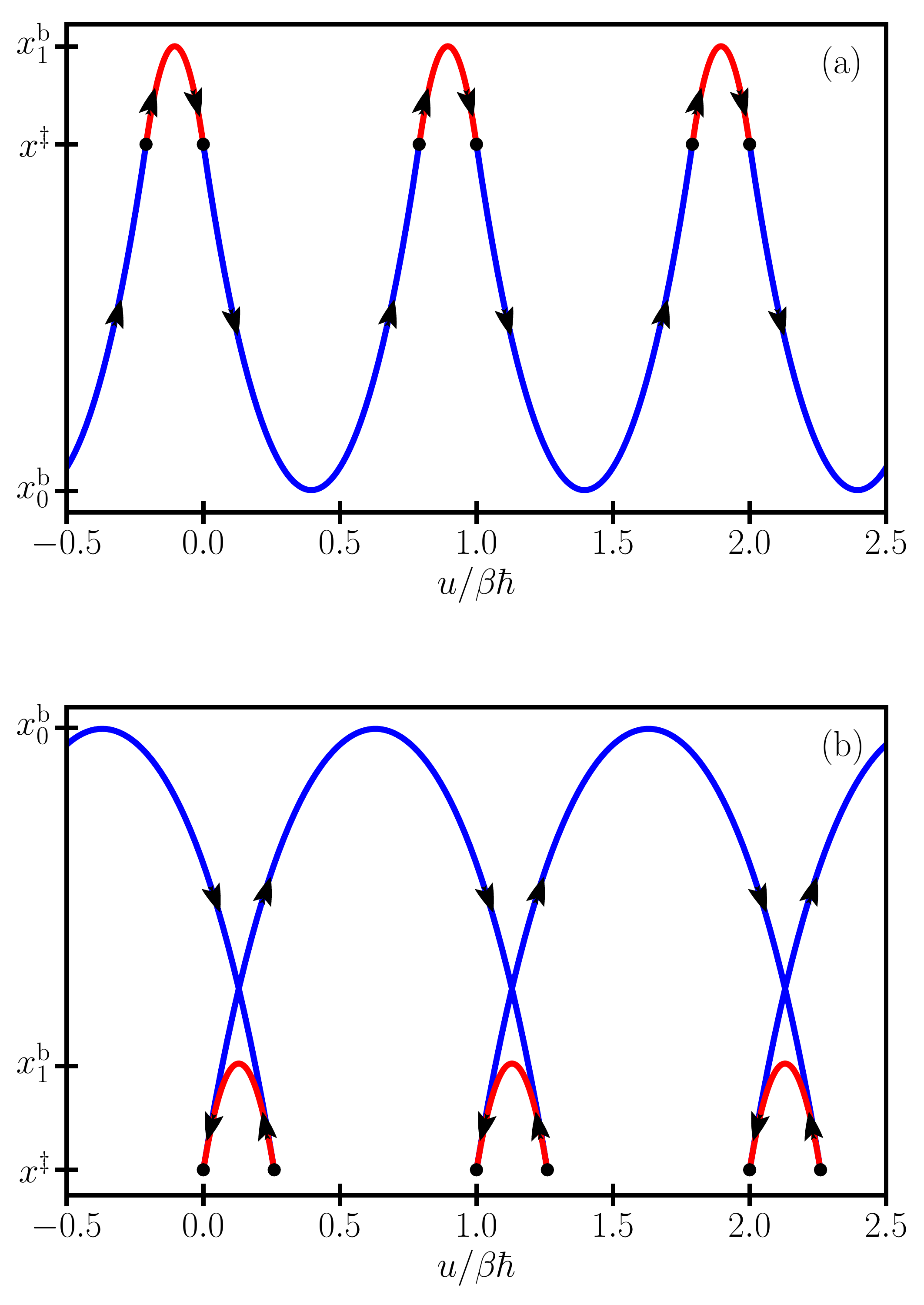}
    \caption{The two trajectories $x_0$ (blue) and $x_1$ (red) which form the instanton 
    are plotted as a function of imaginary time, $u$, in (a) the normal and (b) the inverted regime.
    In both cases the periodicity of the full instanton is $\beta\hbar$ and three cycles are shown.
    In the normal regime the trajectories travel in positive imaginary time and bounce on opposite sides of $x^{\ddagger}$.
    In contrast in the inverted regime both trajectories bounce on the same side and
    in order to have a continuous path, $x_1$ must travel backwards in imaginary time.
    The arrows indicate the direction of particle flow followed by trajectories from their initial point to their final point.
    }
    \label{fig:feynman}
\end{figure}

In this way, we have discovered the form of the instanton appropriate for describing Fermi's golden-rule in the inverted regime.
We did this purely through a consideration of the stationary point of $S$, defined by \eqn{equ:total_action}.
This instanton has a number of important differences when compared with that in the normal regime,
as can be seen from the plots in \fig{feynman} obtained by
joining the two trajectories together to make the instanton orbit.

In the normal regime,
the dynamics are periodic with a periodicity of $\beta\hbar$.
The motion can be described as 
a particle which travels on the reactant state for an imaginary time $\tau_0$,
then suddenly turns into a particle on the product state with the same energy and momentum where it travels for an imaginary time $\tau_1$
before turning back into a reactant-state particle.

In the inverted regime, the instanton formed by joining the two trajectories
is not single valued at certain times.
However, we will still talk about it as an orbit because it remains periodic in imaginary time with periodicity $\beta\hbar$ and has
continuous position and momentum vectors as well as conserving energy along its whole path.
In particular, the energy and momentum are conserved at the two hopping points
even though the path itself takes a sharp change of direction 
when the particle starts to travel in negative imaginary time.
There is a similarity with these pictures and those used to explain the scattering of particles and antiparticles
according to the Feynman--St\"uckelberg interpretation. \cite{Feynman1986antiparticles}
The dynamics of antiparticles are equivalent to those of ordinary particles except that time is reversed,
so that they are found at their final point at an earlier time than their initial point.
\footnote{Imagine watching a movie backwards of a clock falling out of a window.
It still has momentum pointing down, but it travels along a path upwards from the ground to the window and registers a negative elapsed time for this journey.}
Reading Fig.~\ref{fig:feynman}(b) from left to right one sees a single particle travelling on the reactant electronic state.
At imaginary time $u=0$, a new particle/antiparticle pair is created at the hopping point, while the old particle coexists at a different location.
The new particle also travels on the reactant state but the antiparticle on the product state.
At $u=|\tau|$, the antiparticle annihilates with the original reactant particle at the hopping point, $\mat{x}^\ddag$,
leaving only the new reactant particle,
which continues in the role of the original particle when the cycle repeats.

The stationary point which corresponds to the instanton can be classified by an index,
which is defined as the number of negative eigenvalues of the second-derivative matrix of the action.
Knowing this index will be helpful not only for designing an algorithm for finding the instanton
but also
in order to compute the rate in the inverted regime,
for which we need to compute
determinants of 
second derivatives of the action.
In the normal regime, $C_0$ and $C_1$ are always positive because both trajectories are minimum-action paths and $\Sigma$ is negative because the instanton is a single-index saddle point in the space $(\mat{x}',\mat{x}'',\tau)$.
On the other hand, in the inverted regime, the trajectory $\mat{x}_1$ is a maximum-action path and thus all eigenvalues of the matrix $-\pders{S_1}{\mat{x}''}{\mat{x}'}$ are negative.
$C_1$, which is the determinant of this matrix, may thus be positive or negative depending on whether there are an even or an odd number of nuclear degrees of freedom, $D$.
To find the signs of the second derivatives of the total action,
we turn to the classical limit, which was studied in \Ref{GoldenGreens}.
From Eq.~(67) in that paper, one can see that $\Sigma$ has $D+1$ negative eigenvalues and $D$ positive eigenvalues in the inverted regime.
As the instanton moves smoothly from the high-temperature classical limit
to a low-temperature tunnelling mechanism,
there is no reason why the number of negative eigenvalues of $\Sigma$ should change,
so this result holds also in the general case.
Therefore $\Sigma$ will always have the opposite sign from $C_1$, 
ensuring that the square root in the prefactor of \eqn{equ:kinst} remains real.
Hence the same instanton rate expression 
can be uniformly applied across both the normal and inverted regime.
Finally, we conclude that the instanton is a $(D+1)$-index saddle point of $S(\mat{x}',\mat{x}'',\tau)$,
although due to the fact that $\mat{x}_1$ is a maximum-action path,
it will have an even higher index in the space of paths as explained in \secref{RP}.

\subsection{Hamilton--Jacobi formulation}

Up to this point, we have employed the Lagrangian formulation of classical mechanics to define the imaginary-time trajectories.
An alternative approach is provided by the Hamilton--Jacobi formulation,
which uses an action as a function of energy rather than time. \cite{GutzwillerBook}
This leads to further insight into the behaviour of the instanton in the inverted regime.

To derive the energy-dependent action, we start from \eqn{equ:legendre} and write 
$S_1$ in a slightly different way to give
\begin{align}
 \label{equ:jacobi}
    S_1(\mat{x}_\text{i},\mat{x}_\text{f},\tau_1) &= -\int_{\mat{x}_\text{i}}^{\mat{x}_\text{f}} \bar{\mat{p}}_1 \cdot \rmd \mat{x} + E_1\tau_1 \, .
\end{align}
By introducing the antiparticle momentum $\bar{\mat{p}}_1 = -\mat{p}_1$,
we align the direction of the momentum with the particle flow and thereby make the integral strictly positive.
The magnitude of the imaginary-time momentum is $p_n(\mat{x},E_n)= \sqrt{2m[V_n(\mat{x}) - E_n]}$, which is always real and positive in the classically forbidden region where the instanton exists.
These definitions enable us to make the transition from the Lagrange to the Hamilton--Jacobi formalism by defining the abbreviated actions as 
\begin{align}
    W_n(\mat{x}_\text{i},\mat{x}_\text{f},E_n) = \int p_n(\mat{x}(s),E_n) \, \rmd s \, ,
    \label{equ:abbrev_a}
\end{align}
where we have introduced the line element $\rmd s$ defining a metric in the configuration space
such that $(\rmd s)^2 = \rmd\mat{x}\cdot\rmd\mat{x}$.
This gives a line integral along the path, defined in such a way that $\int\rmd s$ is the path length.
Note that $W_n$ is purely positive and is independent of the sign of $\tau_n$ or direction of the trajectories, which has already been accounted for by the sign in \eqn{equ:jacobi}.
This definition is therefore symmetric to an exchange of end-points,
i.e.\ $W_n(\mat{x}_\text{i},\mat{x}_\text{f},E_n) = W_n(\mat{x}_\text{f},\mat{x}_\text{i},E_n)$.
The relations $S_n =  \pm W_n + E_n \tau_n$, where the sign is chosen to match the sign of $\tau_n$,
thus can be viewed as Legendre transforms obeying the conditions $\pder{S_n}{\tau_n} = E_n$ and $\pder{W_n}{E_n} = -|\tau_n|$.
It also follows that $\pder{S_n}{\mat{x}_\text{i}} = \pm \pder{W_n}{\mat{x}_\text{i}}$
and $\pder{S_n}{\mat{x}_\text{f}} = \pm \pder{W_n}{\mat{x}_\text{f}}$.

It is well known that classical trajectories can be defined either as paths which make $S_n$ stationary (known as Hamilton's principle)
or paths which make $W_n$ stationary (known as Maupertuis' principle).
Typically classical trajectories are minimum-action paths, \cite{LandauMechanics}
and in the normal regime, both $S_0$ and $S_1$ are minima with respect to variation of the path.
However, in the inverted regime, the product trajectory will be a maximum of $S_1$.
Trajectories which bounce once have a conjugate point \cite{GutzwillerBook}
and so the associated $W_n$ are single-index saddle points with respect to variation of the path. \cite{GoldenGreens}
There is nothing different in the definition of $W_1$ in the inverted regime, so the product trajectory is also a single-index saddle point.

If we define $W\equiv W(\mat{x}',\mat{x}'',E)$ appropriately
in the normal and inverted regimes,
the total action, \eqn{equ:total_action}, of the instanton can also be written
\begin{equation}
    \label{Legendre}
    S(\mat{x}',\mat{x}'',\tau) = W(\mat{x}',\mat{x}'',E)
    + \beta\hbar E,
\end{equation}
where either $E=E_0=E_1$ is a function of $(\mat{x}',\mat{x}'',\tau)$ via $E_n=\pder{S_n}{\tau_n}$
or $\beta$ is a function of $(\mat{x}',\mat{x}'',E)$ via $\beta\hbar = - \pder{W}{E}$.

In order to clarify the definitions in each regime, we give an overview over the most important equations
\begin{widetext}
\begin{align*}
    &\textbf{Normal regime}   \hspace{0.3\textwidth}  &    &\textbf{Inverted regime}\\
    &0<\tau_0<\beta\hbar; \, 0<\tau_1<\beta\hbar & &\tau_0 > \beta\hbar; \, \tau_1 < 0\\
    &S = S_0(\mat{x}',\mat{x}'',\tau_0) + S_1(\mat{x}'',\mat{x}',\tau_1) & 
    &S = S_0(\mat{x}',\mat{x}'',\tau_0) + S_1(\mat{x}'',\mat{x}',\tau_1)\\
    &S_0(\mat{x}_\text{i},\mat{x}_\text{f},\tau_0) = W_0(\mat{x}_\text{i},\mat{x}_\text{f},E_0) + E_0 \tau_0 &
    &S_0(\mat{x}_\text{i},\mat{x}_\text{f},\tau_0) = W_0(\mat{x}_\text{i},\mat{x}_\text{f},E_0) + E_0 \tau_0\\ 
    &S_1(\mat{x}_\text{i},\mat{x}_\text{f},\tau_1) = W_1(\mat{x}_\text{i},\mat{x}_\text{f},E_1) + E_1 \tau_1 &
    &S_1(\mat{x}_\text{i},\mat{x}_\text{f},\tau_1) = -W_1(\mat{x}_\text{i},\mat{x}_\text{f},E_1) + E_1 \tau_1\\
    &W =  W_0(\mat{x}',\mat{x}'',E) + W_1(\mat{x}'',\mat{x}',E) &
    &W =  W_0(\mat{x}',\mat{x}'',E) - W_1(\mat{x}'',\mat{x}',E)
\end{align*}
\end{widetext}
where the definition of $W$ follows in each case from the required relationship in \eqn{Legendre}.
Differentiating \eqn{Legendre} using the chain rule,
we find $\pder{S}{\mat{x}'}=\pder{W}{\mat{x}'}$ and $\pder{S}{\mat{x}''}=\pder{W}{\mat{x}''}$, which shows that
the instanton, which is a stationary point of $S$,
could also be defined as a stationary point of $W$
with respect to $\mat{x}'$ and $\mat{x}''$.
Either the corresponding temperature can be found for a given $E$ using
$\beta\hbar = - \pder{W}{E}$
or $E$ can be varied until this equation is solved for a given $\beta$. \cite{GoldenRPI}

Let us now check that these definitions are consistent with the one-dimensional schematic shown in Fig.~\ref{fig:instvis}.
In the normal regime it is clear that if we change either $x'$ or $x''$, we increase the path length of one trajectory while simultaneously decreasing the path length of the other.
This thus increases one of the abbreviated actions, $W_n$ 
and decreases the other.
In the normal regime, 
the derivative of the total abbreviated action, $W = W_0 + W_1$, vanishes at the hopping point 
because here
$\pder{W_0}{x'}=\pder{W_0}{x''}=-\pder{W_1}{x'}=-\pder{W_1}{x''} = p^\ddag$,
where $p^\ddag = p_0(x^\ddag,E) = p_1(x^\ddag,E)$.
In the case of the instanton in the inverted regime, changing the positions of the terminal points $x'$ or $x''$ leads to either an elongation or contraction of both trajectories.
In this case, the total abbreviated action has been defined as $W = W_0 - W_1$
and its derivative also vanishes at the hopping point
because here
$\pder{W_0}{x'}=\pder{W_0}{x''}=\pder{W_1}{x'}=\pder{W_1}{x''} = -p^\ddag$.

Furthermore, it is possible to show in this formulation that
in the normal regime,
$\pders{W}{x'}{x'} = \pders{W}{x''}{x''} = 2m[\grad V_0(x^\ddag) - \grad V_1(x^\ddag)]/p^\ddag>0$
whereas in the inverted regime,
$\pders{W}{x'}{x'} = \pders{W}{x''}{x''} = -2m[\grad V_0(x^\ddag) - \grad V_1(x^\ddag)]/p^\ddag<0$,
and in both cases $\pders{W}{x'}{x''}=0$.
Therefore the instanton is defined by 
a minimum of $W(x',x'')$ in the normal regime, but a maximum in the inverted regime.

\subsection{Interpretation of the mechanism}

Finally, we wish to check that the analytically continued instanton formula gives a reasonable
physical model of tunnelling in the inverted regime.
Neglecting the less-important prefactors, the rate is proportional to $\eu{-S/\hbar} = \eu{-\beta E} \eu{-W/\hbar}$,
where we have used \eqn{Legendre} to separate the exponential into two terms.
In a similar way to the standard instanton approach, \cite{Perspective}
we can interpret the 
first term as the probability of
the system acquiring energy $E$ from
thermal fluctuations 
and the
second term as being the probability of tunnelling at this energy.

In order to make this interpretation,
we require that $W$ is positive, ensuring that the tunnelling probability is bounded by 1.
Noting that $W_n \ge 0$,
this is clearly the case in the normal regime, where $W=W_0+W_1$.
However, in the inverted regime, where $W=W_0-W_1$, we will need to show that $W_0 \ge W_1$.
This can be justified, 
at least for a system similar to the schematic in \fig{instvis}(b),
using the fact that $V_0(\mat{x}) \ge V_1(\mat{x})$ and thus $p_0(\mat{x},E_0) \ge p_1(\mat{x},E_1)$ for any point $\mat{x}$ along the instanton.
Noting also that the path length on the reactant state is longer,
it is easy to see from the definition of $W_n$ in \eqn{equ:abbrev_a} that $W_0 \ge W_1$ in this case.

As a corollary to the proof that $W \ge 0$,
by taking the potential to be zero at the reactants, such that 
$V_0(\mat{x}^{(0)}_\text{min})$,
which ensures that $E\ge0$ for the instanton, it follows that $S\ge0$.
This suggests that, at least within the approximation of neglecting the prefactors,
the fastest rate will be found when $S=0$, which is the activationless regime.
There is a barrier to reaction (relative to the reactant minimum)
in both the normal and inverted regime, which gives a positive action and thus smaller rate.
One thus recovers the Marcus turnover picture.
However, because in the inverted regime, $W$ is a difference rather than a sum of two contributions,
it will typically be smaller than in the normal regime,
leading to a larger tunnelling effect.

For the standard instanton approach in the Born--Oppenheimer limit,
it is known that the $\eu{-W/\hbar}$ factor can be derived from one-dimensional WKB theory
as the probability of reaction at a given energy, $E$. \cite{Miller1975semiclassical,Perspective}
WKB theory has also been applied to the nonadiabatic transfer problem
and the result is known as the Zhu--Nakamura formula. \cite{Zhu1995ZN,NakamuraNonadiabatic}
In the diabatic limit their result (written in our notation) for the inverted regime is $\eu{-(W_0-W_1)/\hbar}$,
where the $W_n$ values are calculated along paths from the crossing point to the turning points and back again.
This is identical to the factor found here from analytic continuation of instanton theory and
confirms that our new formula provides the correct physical description of the reaction
according to the mechanistic interpretation suggested in Fig.~\ref{fig:instvis}.

Note that our formula can be applied to general multidimensional systems,
whereas the Zhu--Nakamura formula is a one-dimensional theory. 
Therefore, only with instanton theory will it be possible to study the dominant tunnelling pathways in multidimensional systems,
which in general will involve all degrees of the system,
and will typically not require that
the $\mat{x}_1$ trajectory follows part of the same path as $\mat{x}_0$.

\section{Ring-polymer instanton formulation} 
\label{sec:RP}

\begin{figure}
    \includegraphics[width=8.5cm]{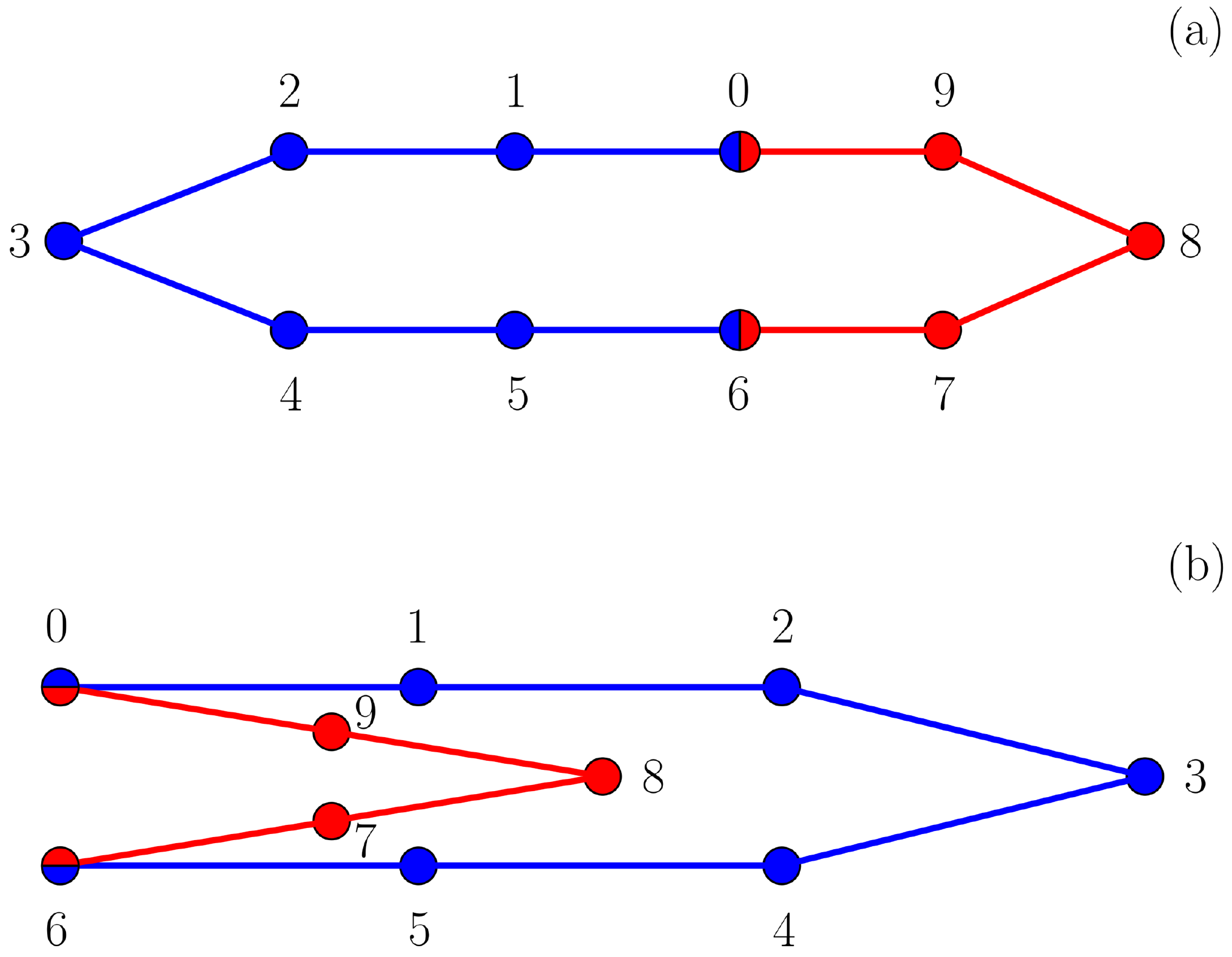}
    \caption{Schematic showing the ring polymer
    corresponding to \eqn{SRP}
    for an example with $N=10$ split into $N_0=6$ and $N_1=4$.
    There is no fundamental difference in the setup of the ring polymer between two regimes,
    but for clarity we show the configurations into which the beads will automatically arrange themselves
    for (a) the normal and (b) the inverted regime.    
    The beads are shown as circles coloured blue if they feel the reactant potential, $V_0$,
    and red if they feel the product potential, $V_1$.  Beads 0 and $N_0$ feel an averaged potential.
    The springs between beads are represented by lines coloured blue for an imaginary-time interval of $\eta_0$ and red for an imaginary-time interval of $\eta_1$.
    In the inverted regime, $\eta_1$ is negative and thus the springs are repulsive.
    }
    \label{fig:Beads}
\end{figure}

In practical implementations of the instanton approach,
we typically 
adopt a ring-polymer formulation. \cite{Schwieters1998diabatic,Andersson2009Hmethane,RPInst,Rommel2011locating,Perspective}
For golden-rule instanton calculations, we follow the approach suggested for the normal regime in \Ref{GoldenRPI}
in which
both paths are discretized into $N_n$ equally spaced imaginary-time intervals
of length $\eta_n = \tau_n/N_n$.
This same approach can be adapted for use in the inverted regime, 
where
$\tau_1$ and therefore $\eta_1$ will be negative.
The resulting $N=N_0+N_1$ beads describe the 
ring polymer
as shown in Fig.~\ref{fig:Beads}.

As previously discussed, the instanton corresponds to a stationary point of the action,
which for a path described by a ring-polymer is given by
\begin{multline}
    S_\text{RP}(\mat{x}^{(0)}, \ldots, \mat{x}^{(N-1)}; \tau) = \\
    \sum_{i=1}^{N_0} \frac{m \|\mat{x}^{(i)} - \mat{x}^{(i-1)}\|^2}{2 \eta_0} 
    + \sum_{i=1}^{N_0 - 1} \eta_0 V_0(\mat{x}^{(i)})\\
    + \sum_{i=N_0 + 1}^{N} \frac{m \|\mat{x}^{(i)} - \mat{x}^{(i-1)}\|^2}{2 \eta_1}
    + \sum_{i=N_0 + 1}^{N - 1} \eta_1 V_1(\mat{x}^{(i)})\\ 
    + \eta_0\frac{V_0(\mat{x}^{(0)}) + V_0(\mat{x}^{(N_0)})}{2}  + \eta_1 \frac{V_1(\mat{x}^{(N_0)}) + V_1(\mat{x}^{(N)})}{2}\, ,
    \label{SRP}
\end{multline}
where cyclic indices are implied such that $\mat{x}^{(0)} \equiv \mat{x}^{(N)}$ in order to form a closed orbit.
This is defined such that in the $N_0,N_1\rightarrow\infty$ limit the value of the ring-polymer action at its stationary point
is equal to the semiclassical instanton action, $S$.%
\footnote{In the activationless regime, the value of $\tau$ at the stationary point will be 0, and thus $\eta_1=0$, which leads to problems in evaluating the ring-polymer action.
However, this causes no problems as in this case, we know that the stationary point has all beads collapsed at the hopping point, which is also the minimum of $V_0$, so no ring-polymer optimization is necessary.
Derivatives of the action can then be evaluated analytically.
}

As in the normal regime,
we define the instanton as a stationary point of $S_\text{RP}$ with respect to 
the bead positions, $\{\mat{x}^{(0)},\dots,\mat{x}^{(N-1)}\}$, and $\tau$ simultaneously.
In order to reduce the number of evaluations of the potential energy and its gradient,
one could alternatively employ a half-ring polymer formalism in a similar way to the standard approach. \cite{GoldenRPI,Andersson2009Hmethane}
Although this is valid for both the normal and inverted regimes,
to simplify the explanations of our findings, we shall not take this extra step here.

In the normal regime, the instanton corresponds to a single-index saddle point.
This is because the instanton is formed of two minimum-action paths,
and $S$ is a maximum only in the $\tau$ variable, i.e.\ $\der[2]{S}{\tau}<0$.
However, 
in the inverted regime the stationary point of the action which we associate with our semiclassical instanton is not 
a single-index saddle point.
In the inverted regime we seek a saddle point with $N_0 D$ positive and $N_1 D + 1$ negative eigenvalues.
This can be understood by first considering the situation where
the values of $\tau$ as well as $\mat{x}^{(0)}$ and $\mat{x}^{(N_0)}$ are fixed at the hopping point.
We would need to minimize the action with respect to the beads $\{\mat{x}^{(1)},\dots,\mat{x}^{(N_0-1)}\}$ and maximize it with respect to the beads $\{\mat{x}^{(N_0+1)},\dots,\mat{x}^{(N-1)}\}$ in order to reproduce the instanton configuration.
Note that this gives perfectly reasonable trajectories, as maximizing the action with respect to the second set of beads, which due to the fact that $\eta_1<0$ have repulsive springs,
is equivalent to minimization with attractive springs.
Then, as explained in \secref{analysis}, the variation of the remaining points gives $D$ negative and $D$ positive eigenvalues, and variation of $\tau$ gives one additional negative eigenvalue.

Starting from an initial guess, we carry out a stationary-point search and thereby simultaneously optimize the bead positions, $\{\mat{x}^{(0)},\dots,\mat{x}^{(N-1)}\}$, and $\tau$.
In the normal regime, \cite{GoldenRPI} we use a standard single-index saddle point optimization algorithm.\cite{WalkingOnPES}
However, 
due to the nature of the instanton in the inverted regime as a higher-index saddle point we have to adapt the optimization scheme slightly.
Finding such higher-index saddle points can sometimes be a very demanding task.
For instance, standard root finding algorithms such as MINPACK's \texttt{hybrd} and \texttt{hybrj} routines (based on a modified Powell method) as implemented for example in scipy can be used.
Note that these approaches locate stationary points of any index and thus may not find the instanton unless a very good initial guess is given.
However, this is made simpler as we exactly know the index of the saddle point we are seeking.
This enables us to use eigenvector-following methods
modified so as to reverse the signs of forces 
corresponding to the first $N_1 D+1$ modes.
\cite{WalesEnergyLandscapes,Doye2002}
One can alternatively modify a single-index saddle-point optimizer such as that introduced in \Ref{WalkingOnPES}
by reversing all but the first of these modes.
This is the approach used to optimize the instantons in \secref{results}.

One might worry that 
the number of higher-index stationary points is often larger than the number of minima and single-index saddle points as was seen in \Ref{Doye2002} for studies of Lennard--Jones clusters. 
However, in our case the stationary points of the action have the direct physical interpretation of two classical trajectories which join together into a periodic orbit.
At least for the simple systems we have studied, it is clear that there is only one
periodic orbit which can exist, and thus only one stationary point of the action.
In fact we ran root-finding algorithms (which optimize to any stationary point regardless of its index) starting from randomly chosen initial conditions and did not find any other stationary points of the action.
We therefore conclude that there is no particular problem in locating ring-polymer instantons in the inverted regime.

In practice, we 
make a sophisticated initial guess of the instanton geometry
using our knowledge that the hopping point is located at the crossing seam and about the general shape of the instanton in the inverted regime.
In addition we can
initiate the optimization at a high temperature, where the instanton is known to be collapsed at the 
minimum-energy crossing point and then systematically cool down the system,
which ensures an excellent initial guess for each step.
In this way the instanton optimization in the inverted regime can be made just as numerically efficient and reliable as in the normal regime.

Once the instanton orbit is found,
the derivatives of the action with respect to $\mat{x}'$, $\mat{x}''$ and $\tau$,
which are required for the prefactor,
can be evaluated in terms of the bead positions, potentials, gradients and Hessians
using the formulas given in the appendix of \Ref{GoldenRPI}.
This allows the rate to be computed directly using the formula in \eqn{equ:kinst}.

\section{Application to model systems}
\label{sec:results}

The analytic-continuation procedure
ensures that we retain the correct behaviour in the high-temperature classical limit
and also that instanton theory continues to give the exact result for a system of two linear potentials. \cite{GoldenGreens}
Here we shall test the numerical implementation for a multidimensional and an anharmonic system
to check that it remains well behaved.
We chose to apply our method to the same two model systems as Lawrence and Manolopoulos in their work on the extrapolation of Wolynes theory into the inverted regime. \cite{Lawrence2018Wolynes}

\subsection{Spin-boson model}
\label{subsec:SB}

The first model system under consideration is the spin-boson model at $T= \SI{300}{\K}$ defined by the potentials
\begin{subequations}
\begin{align}
    V_0(\mat{x}) &= \sum_{j=1}^D \tfrac{1}{2} m \omega_j^2(x_j + \zeta_j)^2 \, ,\\
    V_1(\mat{x}) &= \sum_{j=1}^D \tfrac{1}{2} m \omega_j^2(x_j - \zeta_j)^2 - \varepsilon \, ,
\end{align}
\end{subequations}
where $\zeta_j = c_j/m\omega_j^2$ and
\begin{align}
    \nonumber \omega_j = \omega_{\mathrm{c}} \, \tan\frac{(j - \frac{1}{2})\pi}{2D}& \, , \qquad
    c_j = \sqrt{\frac{m\Lambda}{2D}} \omega_j \, ,
\end{align}
with reorganization energy $\Lambda = \SI{50}{\Calorie\per\mol}$
and characteristic frequency $\omega_{\text{c}} = \SI{500}{\per\cm}$.
This has a
discretized spectral density in $D$ degrees of freedom,
\begin{align}
    J(\omega) &= \frac{\pi}{2} \sum_{j=1}^{D} \frac{c_j^2}{m\omega_j} \delta(\omega - \omega_j) \, ,
\end{align}
which reproduces a Debye spectral density in the $D\rightarrow\infty$ limit. \cite{Wang2003RuRu,Berkelbach2012hybrid}

The exact quantum golden-rule rate for this system with constant diabatic coupling, $\Delta$, can be calculated by numerical integration of the flux correlation function\cite{Weiss} 
\begin{align}
    \label{equ:kex}
    k_{\mathrm{QM}} &= \frac{\Delta^2}{\hbar^2} \int_{-\infty - \mathrm{i}\tau}^{\infty - \mathrm{i}\tau} \mathrm{e}^{-\phi(t)/\hbar} \,\rmd t \, ,
\end{align}
with
\begin{align}
    \phi(t) &= -\mathrm{i}\varepsilon t + \frac{4}{\pi} \int \frac{J(\omega)}{\omega^2} \left[   \frac{1 - \cos{\omega t}}{\tanh{\half \beta \hbar \omega}} + \mathrm{i}\, \sin{\omega t} \right] \rmd \omega \, ,
\end{align}
where the rate is independent of $\tau$, which can therefore be chosen in order to get a faster convergence of the time-integral.

A commonly used approach \cite{Bader1990golden} is to perform a stationary-phase approximation to the rate expression in \eqn{equ:kex} about the point $t = -\mathrm{i}\tau$ such that $\phi'(-\mathrm{i}\tau) = 0$
\begin{align}
    k_{\mathrm{SP}} = \frac{\Delta^2}{\hbar^2} \sqrt{\frac{2\pi\hbar}{\phi''(-\mathrm{i}\tau)}} \, \mathrm{e}^{-\phi(-\mathrm{i}\tau)/\hbar} \, ,
    \label{equ:statphase}
\end{align}
where the primes denote derivatives with respect to $t$.

In the case of the spin-boson model the closed-form expressions for the actions of classical trajectories on each surface are known\cite{Kleinert,Feynman} and the instanton rate can be calculated analytically.
\cite{GoldenGreens,Cao1997nonadiabatic}
The derivation, starting from \eqn{equ:kinst2}, in the inverted regime is completely analogous.
Note that the semiclassical partition function is exact for this harmonic system, i.e.\ $Z_0=\prod_{j=1}^D [2\sinh{\beta\hbar\omega_j/2}]^{-1}$.
The action at the stationary point in the spatial coordinates is given by 
\begin{equation}
    \label{Sspinboson}
    S(\tau) = -\varepsilon\tau + \sum_{j=1}^{D} 2m\omega_j \zeta_j^2 \left[ \frac{1 - \cosh{\omega_j\tau}}{\tanh{\thalf\beta\hbar\omega_j}} + \sinh{\omega_j\tau} \right],
\end{equation}
which holds for both positive and negative $\tau$.
In the case of the spin-boson model, the prefactor in \eqn{equ:kinst2} can be shown to cancel with the reactant partition function.\cite{GoldenGreens} 
Therefore the rate expression reduces to
\begin{equation}
    k_{\mathrm{SCI}} = \sqrt{2\pi\hbar} \, \frac{\Delta^2}{\hbar^2} \left( - \frac{\rmd^2 S}{\rmd\tau^2} \right)^{-\frac{1}{2}} \mathrm{e}^{-S(\tau)/\hbar} \, ,
\end{equation}
which should be evaluated at a value of $\tau$ found numerically to be the stationary point of the action, \eqn{Sspinboson}.
This expression however coincides exactly with the stationary-phase approximation given in \eqn{equ:statphase},
as we identify $S(\tau)\equiv\phi(-\iu\tau)$.
This therefore shows that, for the spin-boson model, the analytically continued instanton theory is equivalent to the stationary-phase approximation in both the normal and inverted regimes.

In order to demonstrate that the ring-polymer instanton optimization and rate calculation
are numerically stable, we carried out calculations using a general multidimensional algorithm,
which did not take the fact that we could solve the problem analytically into account.
The bath was discretized into $D=12$ degrees of freedom as in \Ref{Lawrence2018Wolynes}.
We present all results
as functions of the driving force, $\varepsilon$, and
compare the computed rate constants with 
those of Marcus theory,
\begin{equation}
    k_{\mathrm{MT}}(\varepsilon) = \frac{\Delta^2}{\hbar} \sqrt{\frac{\pi\beta}{\Lambda}}\mathrm{e}^{-\beta(\Lambda - \varepsilon)^2/4\Lambda} \, .
\end{equation}
Note that Marcus theory is equal to classical TST for this system,
to which instanton theory tends in the classical limit. \cite{GoldenGreens}
The inverted regime is defined by $\varepsilon/\Lambda > 1$.

\begin{figure}
    \includegraphics[width=8.5cm]{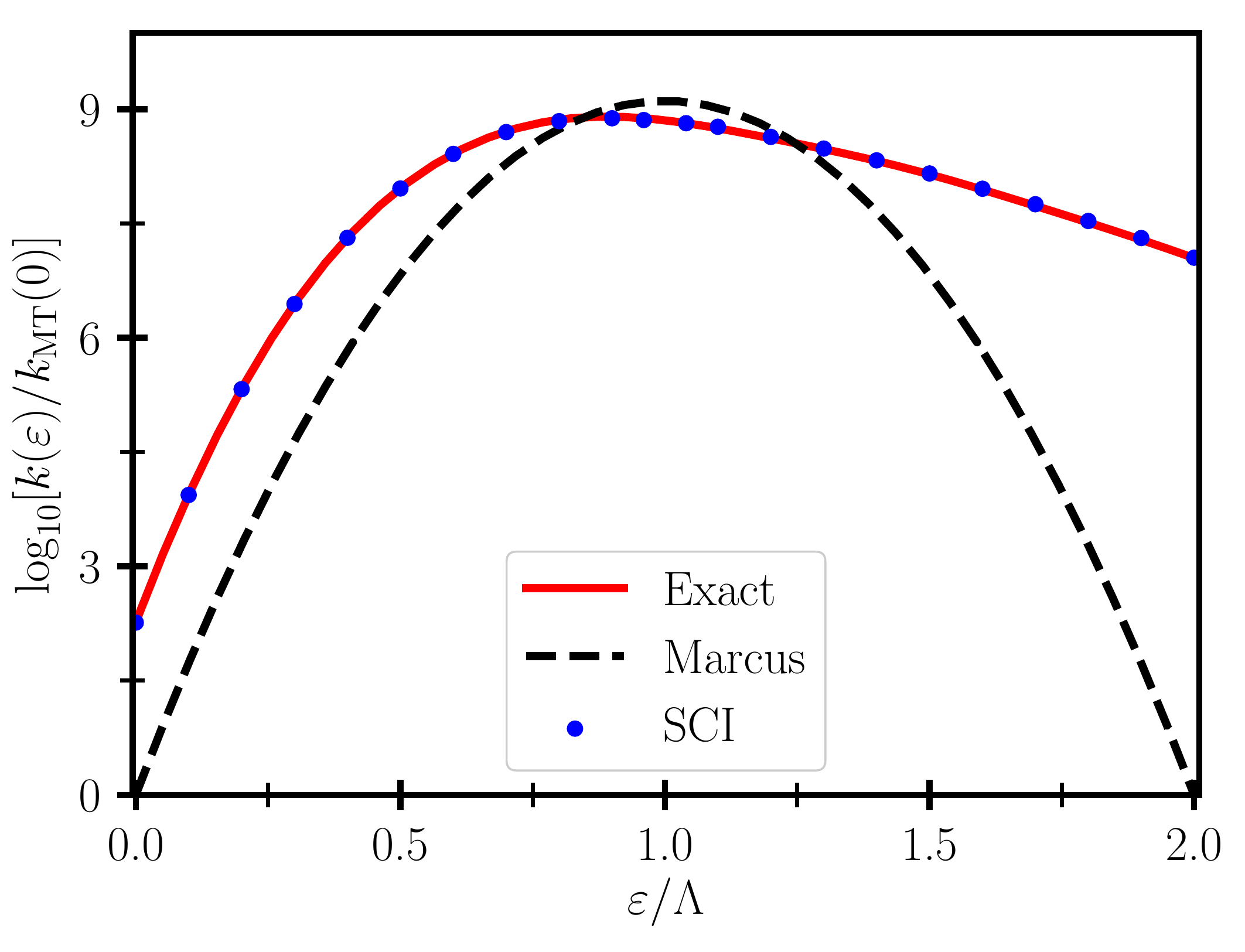}
    \caption{Marcus theory, semiclassical instanton (SCI) and exact quantum rates for a twelve-dimensional spin boson model.
    Results are presented as a function of the driving force relative to the classical Marcus theory rate of the symmetric ($\varepsilon=0$) system.
    }
    \label{fig:SB}
\end{figure}

The semiclassical instanton rates
are depicted in Fig.~\ref{fig:SB}
where they can be compared with the exact result and with Marcus theory.
As expected, instanton theory gives very accurate rates, at least for this system,
as we have explained that it
matches the results obtained by the stationary-phase expression,
which in turn is known to be in excellent agreement with the exact rate for this model.
\footnote{The stationary-phase result deviates from the exact rate by 4\% in the activationless regime,
but the errors for the other cases tested are well below 1\%.}
Furthermore Fig.~\ref{fig:SB} confirms that nuclear quantum effects in the inverted regime can be much larger than those in the normal regime,\cite{Siders1981inverted} causing a dramatic orders-of-magnitude increase in the rate compared with the classical Marcus theory.
It is for this reason that we considered it of particular importance 
to develop the practical instanton approach 
described in this paper.

Table~\ref{table:1} shows how the results converge with the total number of ring-polymer beads, $N$,
for systems in the normal, activationless and inverted regimes.
It can be seen that when the beads are split among the two potentials according to the optimal ratio $N_1/N_0 = |\tau_1|/\tau_0$, the rate converges very quickly.
Even with only $N=128$ beads, all rates are found to be converged to within 2.5\% of the stationary-phase result.
However, in general, $\tau$ is not known prior to the optimization.
Hence we also show the rates optimized with an equal split of the beads, $N_0=N_1$.
Although the instanton rates approach the stationary-phase results slower compared to the rates calculated with optimal ratio, convergence is again reached in all cases.
Consequently a proficient initial guess makes convergence faster but is not required to obtain accurate results.
Furthermore, a simple approach suggests itself in which one could use the optimized $\tau$ value obtained from an instanton search with a low number of beads.
The split of the beads can then be adjusted accordingly for more accurate calculations performed with a larger number of beads.

These results
confirm that 
golden-rule instanton theory is not only as accurate,
but also as efficient in the inverted regime as it is in the normal regime.
In fact convergence to an almost perfect quantitative agreement is achieved
even deep in the inverted regime, where the quantum rate has a $10^7$-fold speed up due to tunnelling.

\begin{table*}
\caption{
Numerical results for the reaction rates of the spin-boson model
(parameters defined in Sec.~\ref{subsec:SB})
computed using various methods
given relative to the Marcus rate for the same system as $\log_{10} [k(\varepsilon)/k_{\mathrm{MT}}(\varepsilon)]$.
The values of $\tau$ given are determined from the calculation of the stationary-phase expression.
We optimize the instanton using two approaches for splitting the $N$ beads into two sets,
one with an optimal bead ratio (to the nearest integers) defined by
$N_1/N_0 = r_{\mathrm{opt}} = |\tau_1|/\tau_0$
(using $\tau$ obtained from the stationary-phase approximation)
and the other with an equal split $N_1/N_0 = r_{1/2} = 1$.
A cell with ``$-$'' indicates failure to find a stationary point with the correct index.
In the limit $N\rightarrow\infty$, the result tends to the stationary-phase approximation (SP), \eqn{equ:statphase}.
Exact rates are calculated by numerically integrating \eqn{equ:kex}.
}
\label{table:1}
\begin{ruledtabular}
\begin{tabular}{L{0.5cm}C{0.1cm} C{0.2cm}C{1.3cm}C{0.1cm} C{1.3cm}C{1.3cm}C{0.1cm} C{1.3cm}C{1.3cm}C{0.1cm} C{1.3cm}C{1.3cm}C{0.1cm} C{1.3cm}C{1.3cm}}
   \multicolumn{2}{r}{$\varepsilon/\Lambda$} & & $0.0$ & & \multicolumn{2}{c}{$0.5$} & &
   \multicolumn{2}{c}{$1.0$} & & \multicolumn{2}{c}{$1.5$} & & \multicolumn{2}{c}{$2.0$} \\
   \multicolumn{2}{r}{\quad$\tau/\beta\hbar$} & & $0.5000$ & & \multicolumn{2}{c}{$0.1589$} & & \multicolumn{2}{c}{$0.0000$} & & \multicolumn{2}{c}{$-0.0430$} & & \multicolumn{2}{c}{$-0.0612$} \\ 
   \cline{4-4} \cline{6-7} \cline{9-10} \cline{12-13} \cline{15-16}
  \multicolumn{2}{l}{$N$} & & $r_{\mathrm{opt}}$ & & $r_{\mathrm{opt}}$ & $r_{1/2}$ & & $r_{\mathrm{opt}}$ & $r_{1/2}$ & & $r_{\mathrm{opt}}$ & $r_{1/2}$ & & $r_{\mathrm{opt}}$ & $r_{1/2}$ \\
 \hline
  \multicolumn{2}{l}{$32$}  & & $2.24$ & & $1.12$ & $1.50$ & & $-0.25$ & $0.60$ & & $1.37$ & $-$ & & $7.00$ & $-$ \\ 
 \multicolumn{2}{l}{$64$} & & $2.26$ & & $1.12$ & $1.26$ & & $-0.26$ & $0.04$ & & $1.31$ & $2.17$ & & $7.01$ & $-$ \\
 \multicolumn{2}{l}{$128$} & & $2.26$ & & $1.13$ & $1.16$ & & $-0.26$ & $-0.18$ & & $1.31$ & $1.47$ & & $7.04$ & $7.19$ \\
 \multicolumn{2}{l}{$256$} & & $2.26$ & & $1.13$ & $1.14$ & & $-0.26$ & $-0.24$ & & $1.32$ & $1.36$ & & $7.05$ & $7.08$ \\
 \multicolumn{2}{l}{$512$} & & $2.26$ & & $1.13$ & $1.13$ & & $-0.26$ & $-0.26$ & & $1.32$ & $1.33$ & & $7.05$ & $7.06$ \\
 \multicolumn{2}{l}{$1024$} & & $2.26$ & & $1.13$ & $1.13$ & & $-0.26$ & $-0.26$ & & $1.32$ & $1.33$ & & $7.05$ & $7.06$\\
 \multicolumn{2}{l}{$2048$}& & $2.26$ & & $1.13$ & $1.13$ & & $-0.26$ & $-0.26$ & & $1.32$ & $1.32$ & & $7.05$ & $7.05$\\
 \hline
 \multicolumn{2}{l}{SP} & & $2.26$ & & \multicolumn{2}{c}{$1.13$} & & \multicolumn{2}{c}{$-0.26$} & & \multicolumn{2}{c}{$1.32$} & & \multicolumn{2}{c}{$7.05$} \\
 \multicolumn{2}{l}{Exact} & & $2.26$ & & \multicolumn{2}{c}{$1.13$} & & \multicolumn{2}{c}{$-0.25$} & & \multicolumn{2}{c}{$1.31$} & & \multicolumn{2}{c}{$7.05$} \\
\end{tabular}
\end{ruledtabular}
\end{table*}

\subsection{Predissociation model}

\begin{figure}
    \includegraphics[width=8.5cm]{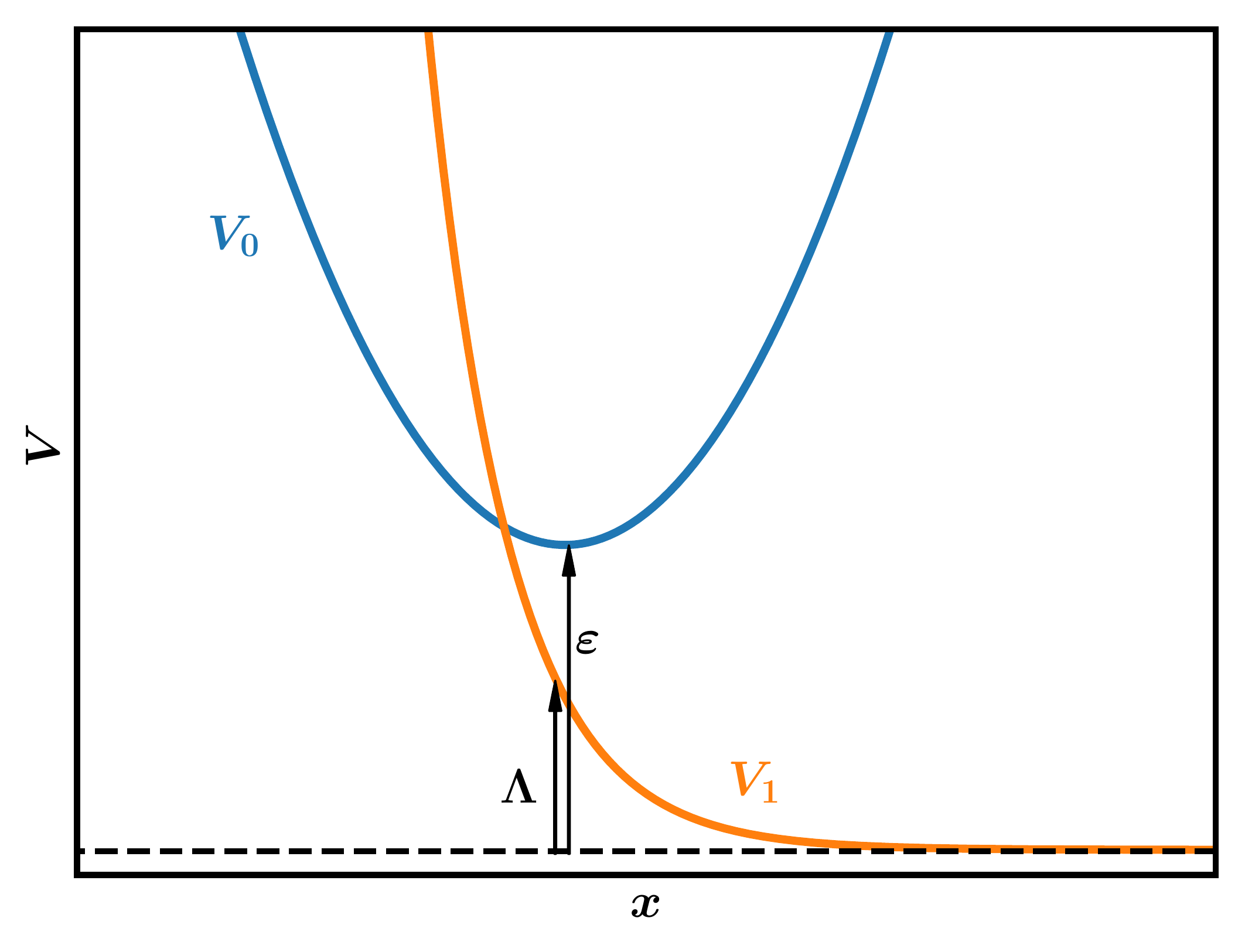}
    \caption{The diabatic potential-energy curves for the one-dimensional predissociation model
    for a particular choice of driving force, $\varepsilon$, in the inverted regime.
    The reorganization energy, $\Lambda$, is also indicated.
    }
    \label{fig:predissociation}
\end{figure}

In this section, we 
show that the approach is not restricted to harmonic systems, but works just as well 
for anharmonic potentials,
which is of course the main advantage of the instanton approach.
We consider the predissociation model previously studied in \Refs{nonoscillatory} and \onlinecite{Lawrence2018Wolynes},
which is not only anharmonic and asymmetric but also contains an unbound state.
The potentials are given as
\begin{subequations}
\label{equ:pd_pot}
\begin{align}
    V_0(x) &= \thalf m\omega^2x^2 \, ,\\
    \label{equ:pd_pot1}
    V_1(x) &= D_{\text{e}} \mathrm{e}^{-2\alpha(x-\zeta)} - \varepsilon \, ,
\end{align}
\end{subequations}
with reduced units $m=1$, $\omega = 1$, $D_{\text{e}} = 2$, $\alpha = 0.2$, $\zeta = 5$, $\beta = 3$ and $\hbar = 1$,
whereas $\varepsilon$ is varied.
Both states are depicted in Fig.~\ref{fig:predissociation} for one particular choice of the driving force, $\varepsilon$.
We present results for a range of values of driving force relative to
the reorganization energy given by $\Lambda=D_{\text{e}}\mathrm{e}^{2\alpha\zeta}$, which is the key parameter for determining the crossover between normal and inverted regimes.
The exact quantum golden-rule rate expressed in the eigenbases of reactant and product states \cite{nonoscillatory} $\{\psi_0^{\lambda}, \psi_1^{\nu}\}$ initialized by a thermal distribution of reactant states is calculated using \eqn{equ:k_qs}.
Just as in the spin-boson example we give all rates relative to the corresponding classical rate.
For the predissociation model, with $\Delta$ taken to be constant, the classical limit is given by the one-dimensional classical golden-rule transition-state theory (cl-TST) rate, 
\eqn{clTST},
where $Z_0^\text{cl} = 1/\beta\hbar\omega$.

\begin{figure}
    \includegraphics[width=8.5cm]{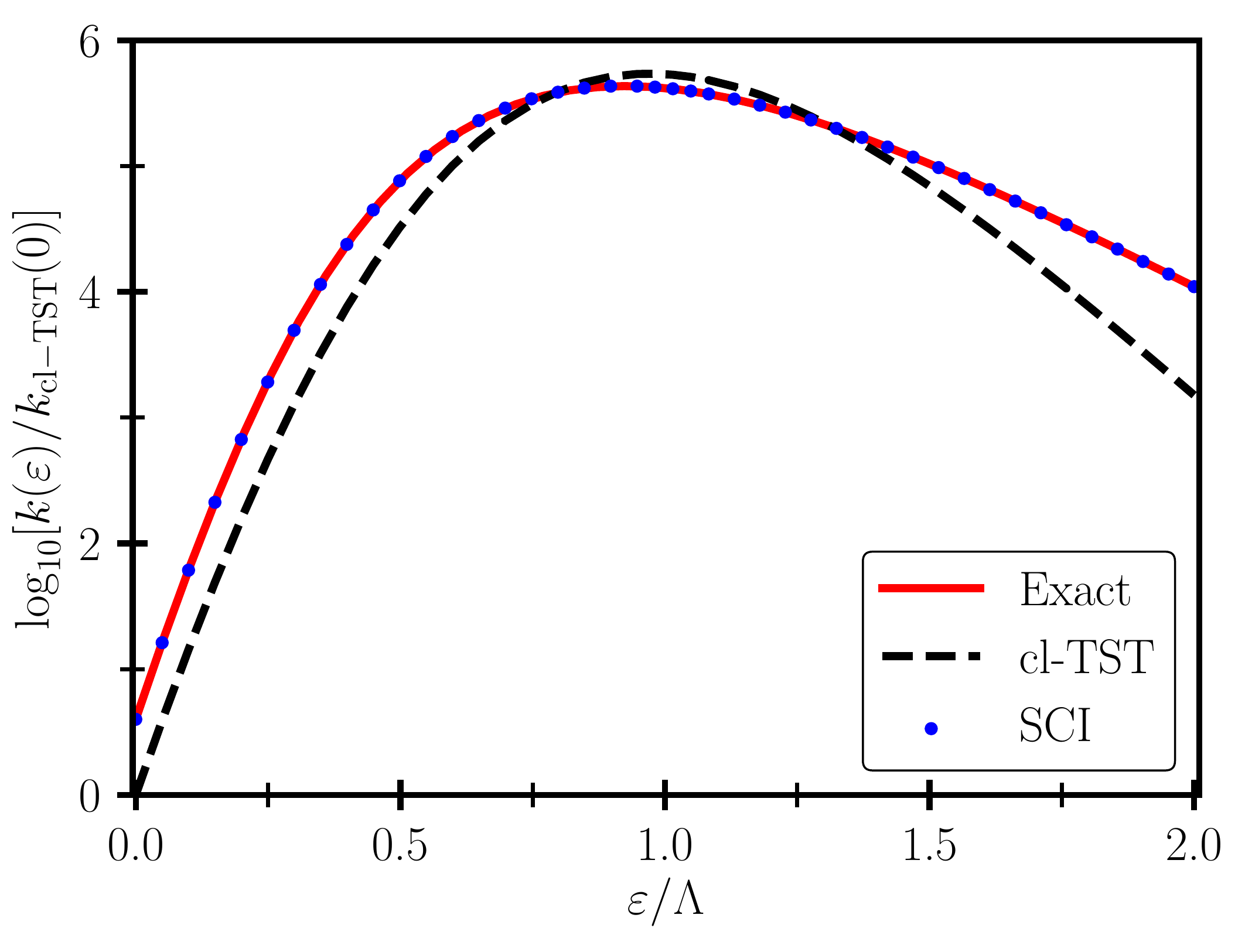}
    \caption{Semiclassical instanton (SCI), exact quantum (QM) and classical TST rates for the predissociation model.
    Results from the various methods are presented as a function of the driving force, $\varepsilon$,
    relative to the classical rate of the $\varepsilon=0$ system.
    }
    \label{fig:PD}
\end{figure}

The results are depicted in Fig.~\ref{fig:PD}, again showing excellent agreement between the exact and instanton rates with a maximal relative error of 0.1\%.
The order-of-magnitude deviation of the classical rate for large $\varepsilon$ emphasizes the remarkable relevance of nuclear tunnelling effects in these systems especially in the inverted regime, which can be almost perfectly captured with our semiclassical instanton approach.
Note that although Lawrence and Manolopoulos were able to achieve similarly accurate results with their approach, \cite{Lawrence2018Wolynes}
it was necessary for them to design a special functional form to treat this dissociative system,
whereas we could apply an identical algorithm to the case of the spin-boson model.

Besides the calculation of electron-transfer rates, we want to stress another interesting possible application of the method, for which the predissociation model provides a simple example.
Instead of artificially shifting two potential-energy surfaces in order to simulate different regimes of electron transfer, the shift between the two surfaces could be caused by a variable external field. 
For instance, we consider a system with a ground electronic state $\ket{0}$
which is uncoupled to an excited electronic state $\ket{1}$
and these potential-energy surfaces may be well separated in energy.
We can then study the situation
of the interaction of the uncoupled system with a light field with continuous-wave frequency $\omega_{\mathrm{ex}}$ 
and interpret the golden-rule result as a photodissociation spectrum.
The two electronic states will now be coupled by the electric dipole operator $\mu(\hat{\mat{x}})$
instead of the nonadiabatic coupling, $\Delta(\hat{\mat{x}})$.
Hence, the golden-rule limit is equivalent to a linear response treatment of the weak-field interaction.

The reason why we can use the instanton method for this problem is because, like the rate, it is also described by 
Fermi's golden rule.
The simple connection between the definitions of the rate defined by \eqn{equ:k_qs} and the total photodissociation cross section in the weak-coupling limit initialized by a thermal equilibrium distribution becomes apparent when looking at the formula for the total photodissociation cross section starting from a thermal equilibrium distribution\cite{Tannor,Schinke_1993,Manolopoulos1992}
\begin{multline}
    \sigma_{\mathrm{tot}} (\omega_{\mathrm{ex}}) = \frac{\pi\omega_{\mathrm{ex}}}{\epsilon_0 c}
    \sum_{\lambda} \frac{\mathrm{e}^{-\beta E_0^{\lambda}}}{Z_0}\\ 
    \times \int |\mu_{\lambda\nu}|^2 \delta(E_0^{\lambda} + \hbar\omega_{\mathrm{ex}} - E_1^{\nu}) \, \rmd E_1^{\nu} \, ,
\end{multline}
where $c$ is the speed of light, $\epsilon_0$ is the vacuum permittivity
and $\mu_{\lambda\nu} = \int \psi_0^{\lambda}(\mat{x})^* \mu(\mat{x}) \psi_1^{\nu}(\mat{x}) \,\rmd \mat{x}$. Note that in our example of a scattering excited state the energies $E_1^{\nu}$ are continuous. Therefore we have replaced the sum in \eqn{equ:k_qs} by an integral and used  energy-normalized continuum wave functions $\psi_1^{\nu}$.
Here we shall assume the transition dipole moment to be constant, also known as the Condon approximation.
Hence the total cross section is directly related to golden-rule rate theory by
\begin{equation}
    \sigma_{\mathrm{tot}}(\omega_\text{ex}) = \frac{\hbar\omega_{\mathrm{ex}}}{2\epsilon_0 c} k_\text{QM}(\hbar\omega_\text{ex}) \, ,
    \label{equ:sofk}
\end{equation}
where the rate constant, $k_\text{QM}(\hbar\omega_\text{ex})$,
is computed 
with 
the reactant potential shifted by the photon energy, i.e.\ $V_0(\mat{x}) \rightarrow V_0(\mat{x}) + \hbar\omega_\text{ex}$
and 
$\Delta$ is replaced by $\mu$. 
The rate thus depends on the photon frequency in the sense that it shifts the potential-energy surfaces relative to each other.
Similar expressions can be given for spontaneous or stimulated emission. \cite{Tannor}

\begin{figure}
    \includegraphics[width=8.5cm]{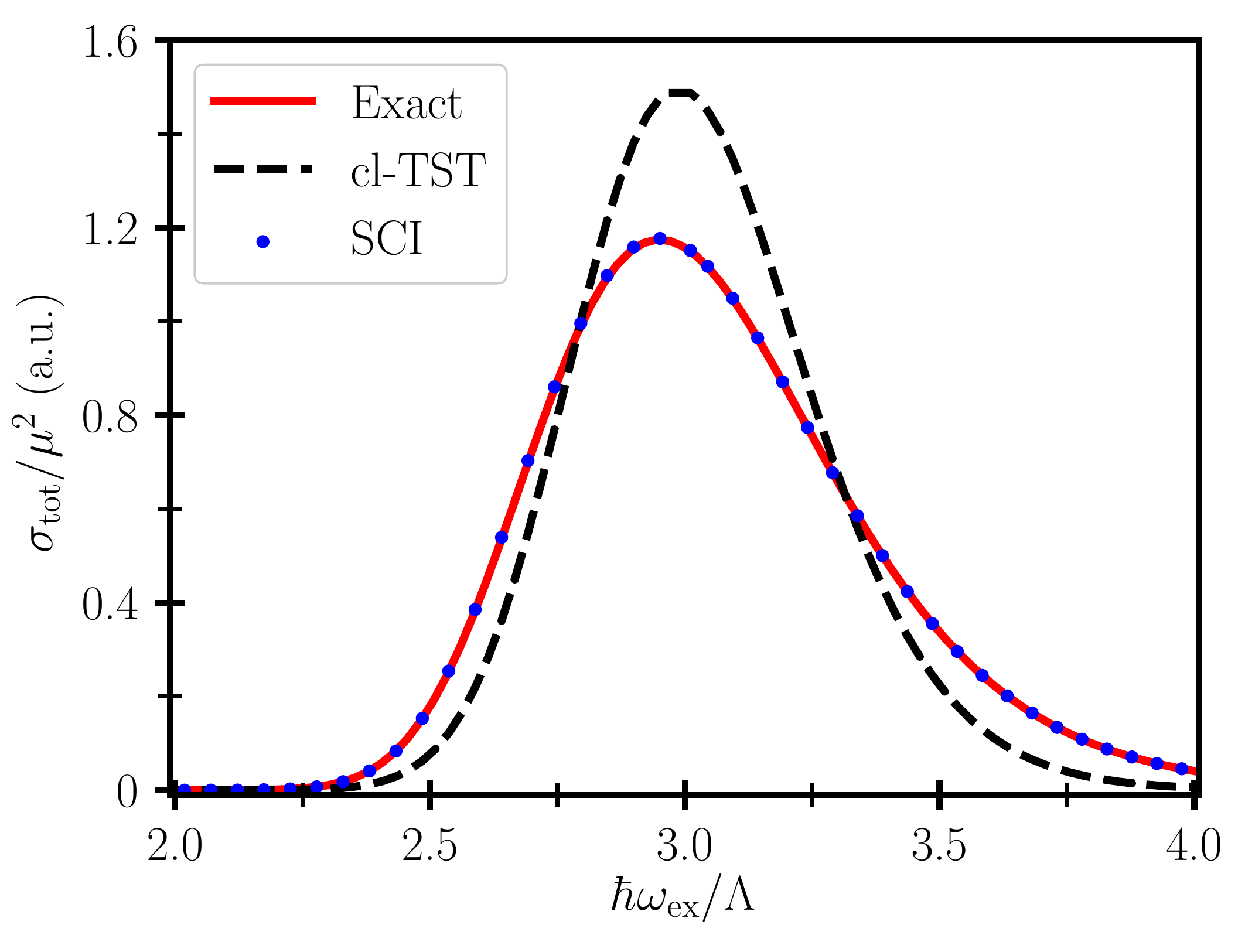}
    \caption{Total photodissociation cross sections for the predissociation model obtained by semiclassical instanton (SCI), exact quantum (QM) and classical TST calculations.
	 The dissociative excited-state potential, $V_1$, is given an asymptotic energy of $-\varepsilon=2\Lambda$
	 and is coupled to the ground-state potential, $V_0$, by a continuous-wave weak field with 
	 frequency $\omega_{\mathrm{ex}}$.
	 }
    \label{fig:sigma}
\end{figure}

Replacing the exact rate in \eqn{equ:sofk} by the instanton or the classical TST rate,
we obtain the various approximate simulated spectra shown in Fig.~\ref{fig:sigma}.
In this case, we used 
the predissociation model [\eqs{equ:pd_pot}] with a fixed value of $\varepsilon=-2\Lambda$,
which then describes the typical case of an excited electronic-state potential $V_1$ 
high above the ground-state potential $V_0$.
The deviation of the classical cross sections again illustrates the sizeable effect of quantum nuclear effects, this time on the line-shape of optical spectra.
On the other hand, semiclassical instanton theory reaches graphical accuracy with the exact result.

In order to showcase what our method can contribute to the description of spectra
it is worth making a comparison with
standard approaches used in quantum chemistry. 
The simplest and probably the most common method
is to calculate the vertical excitation energy from the ground state minimum, which corresponds to $\Lambda-\varepsilon$ in our model (specifically $3\Lambda$ according to our choice of parameters).
This gives a single peak in the spectrum at $\hbar\omega_\text{ex} = \Lambda-\varepsilon$ and
with this approach one completely disregards the statistics (or dynamics) of both states.
This method can be improved by 
assuming a classical Boltzmann distribution in the ground state, which can, for example, be sampled by molecular dynamics simulations.
By calculating vertical excitation energies from different sampled configurations the natural width of the absorption bands can be revealed.
This corresponds to our classical TST calculations.
The instanton method improves this description 
by including quantum nuclear effects for both the ground and excited state.
As can be seen in Fig.~\ref{fig:sigma} the absorption maximum in both the quantum and semiclassical calculations is slightly shifted to lower energies compared to the classical result. This shift ($\approx 0.03\Lambda$) is a direct consequence of the reactant potential's zero-point energy, which is not accounted for in the classical calculations. Furthermore, due to tunnelling in the normal regime the absorption band exhibits  an earlier onset, whereas the equivalent process in the inverted regime causes a slower decay of the band.
The speed up of the transition rate induced by quantum tunnelling therefore directly translates into adding longer tails to both sides of the absorption spectrum.

We do however ignore the real-time dynamics within the wells
such that 
our method thus probes only the time-scale of the fastest decay of the wave packet which imprints itself in the broadest features of the spectrum.
These features form an envelope of constraint on the spectrum that will not be changed by any other dynamics
leading to vibrational and rotational fine structure. \cite{HellerBook}
Therefore, although we shall not be able to describe vibrational progressions with this approach,
we expect to predict the correct envelope of the spectrum.
We note that this example of transition to a dissociative state is thus a particularly favourable case for us.

\section{Conclusions}

We have extended the semiclassical instanton method
for calculating Fermi's golden rule into the Marcus inverted regime.
It can be applied to multidimensional and anharmonic systems and gives a good approximation to the exact result even when nuclear quantum effects are important.
The theory reduces to classical golden-rule TST in the high-temperature limit and hence Marcus theory when the potentials are harmonic,
is exact for a system of two linear potentials,
and
is identical to the stationary-phase approximation in the case of the spin-boson model.

The main difference between the normal and inverted regimes
is the form of the instanton periodic orbit,
although in both cases it is
defined by the stationary point of the total action formed by joining two imaginary-time trajectories together.
In the normal regime, both trajectories are minimum-action paths which travel in positive imaginary time on the reactant or product potential-energy surfaces.
However, in the inverted regime, the product trajectory is a maximum-action path which travels in negative imaginary time.
In both regimes, the energy and momentum are conserved when hopping from one state to the other,
which occurs at a point where the two diabatic potentials are degenerate.

In order to locate the inverted-regime instanton within the ring-polymer formalism,
we search for a high-index saddle point of the total action.
We show that by using the knowledge we have about the expected number of negative eigenvalues 
as well as the approximate shape and location of the two trajectories, the algorithm can be made just as efficient as in the normal regime.
Therefore 
this approach can be used
to calculate reaction rates across the whole range of electron-transfer reactions
or for simulating spectral line shapes.

In contrast to closed-form expressions for the rate, \cite{NakamuraNonadiabatic,Jang2006ET}
which effectively require a \emph{global} harmonic approximation for the potential-energy surfaces,
the instanton approach locates the instanton without making any assumptions
and takes only a \emph{local} harmonic approximation about the tunnelling path.
All one has to provide to the algorithm
are functions returning the potentials, gradients and Hessians on the two diabatic potential-energy surfaces for a given nuclear configuration.
Additionally it only requires this knowledge about a rather small region 
located around the crossing point.
This is another reason for the computational efficiency of the method and makes it conceptually easily applicable to 
molecular systems,
even in conjuncture with on-the-fly \textit{ab-initio} electronic-structure codes.
For further enhancements to the efficiency, machine-learning approaches could be applied. \cite{GPR}

Apart from the excellent agreement with exact quantum rates for the model systems studied in this work, one of the main advantages of the method from a qualitative perspective is that it provides direct mechanistic insight into the reaction under investigation.
In this respect, our method appealingly stands out from alternative approaches
which effectively extrapolate the data collected in the normal regime. \cite{Lawrence2018Wolynes}
The instanton orbit can be interpreted as the `dominant tunnelling pathway'
and identifies which nuclear modes are involved in the reaction.
In cases where there are competing reactions, the instanton approach will identify the dominant mechanism.
Comparison to the classical (Marcus or golden-rule TST) rate allows easy identification of the role of quantum nuclear effects (tunnelling and zero-point energy),
which are expected to be particularly important in the inverted regime.
Kinetic isotope effects can also be easily predicted, which often provide the easiest connection to experimental data. \cite{MUSTreview}

A limitation of all instanton methods is that the semiclassical approximation is not valid for
liquid systems.
Nonetheless, a good understanding of the instanton paths has helped in the development of a number of methods based on path-integral sampling which are applicable to reactions in solution. \cite{GRQTST,GRQTST2,QInst}
We hope that the information obtained on the instanton in this work
will help derive novel path-integral-based rate theories which can describe the inverted regime more rigorously.

The method described in this work is, however, well suited to be applied to complex chemical reactions
in the gas-phase, in clusters, on surfaces and in solids.
A wide range of processes involving a transition between two electronic states
can be studied in this way,
so long as the coupling falls in the golden-rule regime.
This encompasses not only electron-transfer reactions, but also spectroscopy, intersystem crossing and electrochemical processes.
Showing the capability of the method in such applications will be an integral part of future work.  

\section*{Acknowledgements}
This work was financially supported by the Swiss National Science Foundation through SNSF Project 175696.

%

\end{document}